\documentclass{article}

\usepackage{arxiv}

\usepackage[utf8]{inputenc} % allow utf-8 input
\usepackage[T1]{fontenc}    % use 8-bit T1 fonts
\usepackage[english]{babel}
\usepackage{hyperref}       % hyperlinks
\usepackage{url}            % simple URL typesetting
\usepackage[square,numbers,sort&compress]{natbib}
\usepackage{booktabs}       % professional-quality tables
\usepackage{nicefrac}       % compact symbols for 1/2, etc.
\usepackage{microtype}      % microtypography
\usepackage{graphicx}
\graphicspath{ {./images/} }
\usepackage{bm}
\usepackage{float}
\usepackage{csquotes}
\MakeOuterQuote{"}
\usepackage{amsmath,amsfonts,amsthm,amsopn,amssymb}
\usepackage{mathrsfs}
\usepackage{subfig}
\usepackage{xcolor}

\title{Finding the Underlying Viscoelastic Constitutive Equation via Universal Differential Equations and Differentiable Physics}

\author{
 Elias C. Rodrigues \\
  School of Applied Mathematics\\
  Getulio Vargas Foundation\\
  Rio de Janeiro, RJ, Brazil \\
  \texttt{elias.rodrigues@fgv.br} \\  
 \And
  Roney L. Thompson \\
  COPPE\\
  Federal University of Rio de Janeiro\\
  Rio de Janeiro, RJ, Brazil \\
  \texttt{rthompson@mecanica.coppe.ufrj.br} \\
  \And
 Dário A.B. Oliveira \\
  School of Applied Mathematics\\
  Getulio Vargas Foundation\\
  Rio de Janeiro, RJ, Brazil \\
  \texttt{dario.oliveira@fgv.br} \\
  \And
  Roberto F. Ausas\\
  Institute of Mathematics and Computer Science\\
  University of São Paulo\\
  São Carlos, SP, Brazil \\
  \texttt{rfausas@icmc.usp.br} \\  
}

\begin{document}
\maketitle
\begin{abstract}
This research employs Universal Differential Equations (UDEs) alongside differentiable physics to model viscoelastic fluids, merging conventional differential equations, neural networks and numerical methods to reconstruct missing terms in constitutive models. This study focuses on analyzing four viscoelastic models: Upper Convected Maxwell (UCM), Johnson-Segalman, Giesekus, and Exponential Phan-Thien-Tanner (ePTT), through the use of synthetic datasets. The methodology was tested across different experimental conditions, including oscillatory and startup flows. While the UDE framework effectively predicts shear and normal stresses for most models, it demonstrates some limitations when applied to the ePTT model. The findings underscore the potential of UDEs in fluid mechanics while identifying critical areas for methodological improvement. Also, a model distillation approach was employed to extract simplified models from complex ones, emphasizing the versatility and robustness of UDEs in rheological modeling.
\end{abstract}

\section{Introduction}
\label{Intro}

Constitutive models are a fundamental part of the governing equations associated with the evolution of a generic material. Basic conservation laws are insufficient as they result in an undetermined system with more unknowns than equations. Hence, new equations must come into play to provide the necessary matching. Constitutive models provide additional relationships by defining the behavior of specific material classes, linking forces to kinematics, or, more precisely, stress to deformation and their rates.  Determining the appropriate constitutive model and its associated properties is essential for characterizing a given material. Understanding and describing material behavior lies at the core of Rheology.

Understanding the behavior of non-Newtonian fluids can be achieved through either a phenomenological approach or by observing fundamental principles. The former focuses on analyzing rheometric data and creating an equation to depict the observed macroscopic behavior. For the latter, the model is constructed directly from physical principles without relying on any empirical data \cite{Brader2010}. This can be achieved in different ways. Such as making assumptions about the material’s microstructure and performing an upscaling procedure, analyzing how the material stores and dissipates energy, or using mechanical analogs and extending the framework to three-dimensional tensors. As usual, each strategy has its advantages and drawbacks. Nonetheless, the macroscopic phenomenological method based on continuum mechanics is commonly used for tackling industrial challenges due to its straightforward interpretation and computational simplicity in predicting flow in complex geometries.

Although non-Newtonian phenomenological models are widely used, they can be challenging to develop, as they must be precise enough to cover the different deformation processes encountered in practice while describing the rheology of materials exhibiting processes on various time scales. Rheologists often face a challenging endeavor in deciphering data from various rheological tests to understand the correlation between stress $\bm\sigma$, and strain rate $\bm{\dot\gamma}$, in non-Newtonian fluids, which is influenced by factors such as microstructural and multiscale characteristics, elasticity, plasticity, and time dependency. The model's scope is restricted due to biases in experiments, the extensive range of experiments required for its development, and essential assumptions related to the continuum mechanics framework. In this context, there is potential for additional investigation to refine current equations and experiment with innovative techniques for formulating constitutive equations.

Constitutive rheological models inherently contain errors regardless of the modeling methodology used, especially in complex systems, as they are based on idealizations reflected on the assumptions adopted during the equation development phase. For instance, rheometry is based on imposing simple kinematics conditions on a material, while a complex flow exhibits drastically different behavior. The limited understanding of the specific phenomenon under investigation presents a considerable challenge for some industrial settings in obtaining accurate solutions from the chosen constitutive model.

Recently, the field of Machine Learning (ML) \cite{Jordan2015} has received considerable attention from academia and industry. More specifically, the ML discipline, called Scientific Machine Learning (SciML), employs prior knowledge derived from observational, empirical, physical and mathematical understanding of a phenomenon to bias the machine learning algorithm for solutions that are consistent with scientific knowledge \cite{Karniadakis2021, Brown2024, Iwema2024} and is the focus of the present work.

\emph{Hybrid modeling} is seen as a feasible alternative to overcome the challenges of modeling constitutive equations. This strategy involves some pre-existing knowledge about the problem (e.g., an initial differential equation), where models are developed based on the physical/mathematical description of the phenomenon of interest, but also includes a data-driven method that incorporates data in the same framework. This second step is derived by identifying the inherent patterns in the experimental data and leveraging these data through learning algorithms. This approach aims to mitigate epistemic uncertainty, characterized by incomplete knowledge of the phenomenon being modeled, such that the machine learning step rectifies the discrepancies originating from the poor predictive capabilities of the elementary model \cite{Zendehboudi2018}.

Most studies in the field of SciML and constitutive models focus on solid mechanics, particularly those displaying viscoelastic \cite{Xu2021,Tac2023,Asad2023}, hyperelastic \cite{Flaschel2021,Joshi2022}, or plastic behavior \cite{Haghighat2023,Roy2023,Wang2023}. An opportunity has emerged within constitutive models for fluids, presenting a pathway for exploration and potential development.

A noticeable trend within the rheology community is the growing interest in the emerging SciML technique of Physics-Informed Neural Networks (PINNs) \cite{Raissi2019, Karniadakis2021, Cuomo2022}, as indicated by some studies on the topic of non-Newtonian fluids. For instance, research involving a Rheology‑Informed Neural Networks (RhINNs) for forward and inverse metamodelling of complex fluids \cite{Mahmoudabadbozchelou2021b}; study of a multifidelity neural network (MFNN) architecture for data-driven constitutive metamodeling \cite{Mahmoudabadbozchelou2021a}; investigation of Rheology-informed graph neural networks (RhiGNets) that are capable of learning the hidden rheology of a complex fluid through a limited number of experiments \cite{Mahmoudabadbozchelou2022}; examination of RhINN that enables robust constitutive model selection based on available experimental data \cite{Saadat2022}; analysis of fractional RhINNs used to recover the fractional derivative orders of fractional viscoelastic constitutive models \cite{Dabiri2023}; developing of data-driven rheological characterization of stress buildup and relaxation in thermal greases \cite{Nagrani2023}; a new PINN framework named ViscoelasticNet uses the velocity flow field to choose the best constitutive model for the fluid and to understand the related stress and pressure fields  \cite{Thakur2024}.

The research on PINNs is still in its early stages, just beginning to unfold. In light of this being a nascent field, there are limitations to deal with: notably, it is a limited simulation tool for handling both forward and inverse problems under varying boundary conditions, requiring new training for each specific condition; convergence challenges encountered during residual minimization thus requiring a considerable amount of trial and error to achieve a satisfactory result and therefore demanding significant effort to achieve generalization; the position of training locations affect the results \cite{Wang2020, Wang2022, Rohrhofer2022, Cuomo2022}.

Current research endeavors to examine diverse data-oriented frameworks as a solution to the challenges posed by PINNs. \citeauthor{Jin2023}\cite{Jin2023} introduces the Constitutive Neural Network (ConNN) model, which uses a recurrent neural network structure to understand how stress responses evolve with time. The recurrent units are specifically crafted to mimic the properties of complex fluids, including fading memory, finite elastic deformation, and relaxation spectrum, without making assumptions about the fluid's equation of motion. However, this approach relies solely on data without considering frame invariance and a closed form for the constitutive equation, leading to difficulties in interpretation.

A novel approach known as the SIMPLE (Scattering-Informed Microstructure Prediction during Lagrangian Evolution) method was introduced by \citeauthor{Young2023}\cite{Young2023}, offering a data-driven solution for simulating the behavior of complex fluids in motion based on FFoRM-sSAXS\footnote{Fluidic four-roll mill (FFoRM) - Scanning small-angle X-ray scattering instrumentation (sSAXS).} experiments. Utilizing the Lagrangian trajectory of the fluid within this framework, the authors modeled microstructural aspects and stress evolution during flows by utilizing a Neural ODE equation, which involves solving and training differential equations. Although frame invariance is honoured, the neural networks implemented do not involve physics-based insights, and the method does not construct a closed-form constitutive equation.

The previously cited research on PINNs commences with an already fully defined constitutive equation and aims to fine-tune its parameters based on the provided training data.  However, the constitutive equation remains static and can not adapt to incorporate new parameters, differential operators, tensors or functions.

In a recent study, a novel physics data-driven technique utilizing the Sparse identification of non-linear dynamics (SINDy) method \cite{Brunton2016} was presented by \citeauthor{Mahmoudabadbozchelou2024}  \cite{Mahmoudabadbozchelou2024} for the identification of new non-Newtonian constitutive equations from empirical Carbopol data. An extensive collection of potential function candidates is utilized, and then the Sequential Threshold Ridge Regression (STRidge) algorithm is employed to isolate feasible functions. Despite the promising findings, the approach is not tensorial nor frame indifferent being unable to predict normal stresses. Moreover, the threshold value applied in the regression process impacts the selection of functions, potentially altering the model framework. \cite{Naozuka2022}.

A hybrid approach known as Universal Differential Equations (UDEs) provides a means for the construction of a {new complex model, departing from a simple differential constitutive equation and using available data, with a neural network being inserted directly into the structure of the differential equation. Hence, the data associated with a complex behavior adds new information that is incorporated into the model by the evolution of the network, thus working as a model augmentation procedure. This approach employs a hard constraint since the network is inserted as part of the equation \cite{Rackauckas2021}. Through the integration of adjoint sensitivity analysis and differentiable programming  \cite{Sapienza2024}, the UDEs approach is able to leverage enhanced numerical solvers to compute the solution of a UDE and the gradients of loss functions concerning their inputs, resulting in a fusion of numerical simulations and deep learning known as Differentiable Physics \cite{Liang2019, Ramsundar2021, thuerey2021pbdl}.

The application of universal differential equations extends to disciplines such as civil engineering, mechanics, chemistry, physics, biology, pharmacy, climate science, and others. As an example, the dynamics of a system with non-linear vibration were analyzed by \citeauthor{Lai2021}\cite{Lai2021} through the application of UDEs in structural identification to reduce vibration levels. \citeauthor{Trujillo2023} \cite{Trujillo2023} employed the technique to analyze hysteresis in buildings, demonstrating that the simulation yielded acceptable outcomes despite having limited data. 

When faced with complex phenomena and uncertain physics, UDE-based models can function as a surrogate. \citeauthor{Koch2021}\cite{Koch2021} studied a rotary detonation engine described by a UDE model. The model could effectively differentiate between the scales of the different observed phenomena and could be utilized as a digital twin.

The study by \citeauthor{Jiang2021}\cite{Jiang2021} focused on analyzing a non-Markovian stochastic biochemical kinetics model in gene modeling. Using the EDU, the model's kinetic parameters were approximated. The authors found that their methodology was beneficial in understanding the behaviors of biomolecular processes that were challenging to model with a Markovian approach.

\citeauthor{Keith2021}\cite{Keith2021} introduces UDEs to perform a gravitational waveform inversion strategy that discovers mechanical models of binary black hole systems from gravitational wave measurements. The differential equations derived correspond closely with the numerical calculations of black hole paths. By extrapolating the models in time, they can uncover and incorporate several recognized relativistic effects previously unaddressed in the universal equations.

According to \citeauthor{Nogueira2022}\cite{Nogueira2022}, phenomenological adsorption models have sink/source terms that describe the adsorption equilibrium through a well-known simplified model, so the resulting model is limited by its assumptions. In this sense, they applied the UDE methodology to mitigate the simplifications and use experimental data. The necessary amount of data used to identify the model studied showed that the hybrid model can use a few data to describe the system accurately. Furthermore, the model obtained can describe competitive adsorption more accurately than the Langmuir model. 

\citeauthor{Santana2023}\cite{Santana2023} introduced a methodical machine learning strategy to develop effective hybrid models and identify sorption absorption models within nonlinear advection-diffusion-sorption systems. The research effectively reconstructed the kinetics of sorption absorption, precisely predicting breakthrough curves and emphasizing the possibility of identifying structures of sorption kinetic laws.

The climate modeling field extensively delves into utilizing machine learning methodologies. Some works are being put forward that make use of UDEs. \citeauthor{Ramadhan2020}\cite{Ramadhan2020} employed UDE to model the natural convection of the oceanic boundary layer caused by the surface's loss of buoyancy. By training a UDE with high-resolution explicit models, the authors enhance a simple parameterization of the convective adjustment process to capture fluxes at the base of the mixing layer, which cannot be adequately represented by convective adjustment alone. 

\citeauthor{Bolibar2023}\cite{Bolibar2023} investigated the EDU methodology as a proof of concept in solving a non-linear diffusivity differential equation to learn the creep component of ice flow in a glacier evolution model, enabling the discovery of empirical laws from remote sensing datasets.

One of the few works in Rheology is also presented as a proof of concept by \citeauthor{Lennon2023}\cite{Lennon2023} in the study of complex rheological models with application to flows with abrupt contraction. The authors introduce the concept of RUDEs (Rheological Universal Differential Equations) and suggest that by incorporating differential viscoelastic constitutive equations, these can be seamlessly integrated into current computational fluid dynamics tools. This approach allows for a flexible framework tailored to accommodate new empirical or theoretical insights specific to the material being studied or the application at hand. Nevertheless, a more in-depth investigation into the different viscoelastic models that the technique can accommodate and its limitations remains to be done. The purpose of this present research is to fill in this gap. 

We have organized the rest of this paper into five sections. First, we explain non-Newtonian fluids in more depth, especially viscoelastic fluids, and the type of models to be addressed using the Universal Differential Equation technique. The UDE methodology, including the neural network architecture, is fully outlined in the sequence. Following that, details regarding the dataset employed and the training methodology are presented along with an explanation of the optimization process. The results for each model are then discussed in addition to a proposed surrogate technique presented as Viscoelastic Model Distillation. In closing, a summary and outlook about the methodology are outlined.

\section{Non-Newtonian Fluids}

Non-Newtonian fluids are fluids whose behavior does not adhere to Newton's law of viscosity. Different fluid systems may exhibit departures from Newtonian behavior, generally caused by microelements dispersed on a Newtonian liquid. Hence, microstructured liquidsdisplay a wide range of fluid behaviors, making them a compelling subject of study for various scientific disciplines and industrial applications. Non-Newtonian fluids show unique rheological behaviors that are dependent on the deformation rate, flow-type, temperature, time, and other factors. Such complexities have made non-Newtonian fluids a subject of significant research interest.

Depending on the microstructure nature of the liquid, non-Newtonian fluids exhibit different phenomena such elasticity, plasticity, complex viscous behavior, time dependency, among others. Understanding non-Newtonian fluids involves categorizing their behavior into distinct groups. One such group is the \emph{viscoelastic fluids}, which exhibit viscous and elastic properties. Examples include biological fluids like blood \cite{Beris2021}, cervical mucus \cite{Li2021} and synovial fluid \cite{Fam2007}, along with polymeric materials used in consumer products and industrial processes \cite{Han2007, Sangroniz2023}. Industries such as food processing, cosmetics, pharmaceuticals, and petroleum engineering rely on a comprehensive understanding of viscoelastic fluids to optimize processes and develop innovative products. Rheology is essential to achieving this goal. Rheology concerns the study and description of material deformation. Therefore, Rheology offers a framework for the analysis of the fluid response to different stress field inputs under a variety of conditions, contributing to the understanding of fluid behavior and the construction of constitutive equations.

In fluid flow, the equation of fluid motion is governed by the well-known Cauchy's equation (\ref{eq:Cauchy}): 

Fluid motion is governed by the well-known Cauchy's equation (\ref{eq:Cauchy}), given by

\begin{equation}
	\rho\left(\dfrac{\partial\bm{v}}{\partial t} + \bm{v}\boldsymbol{\cdot}\bm\nabla{\bm{v}}\right)=\bm\nabla\boldsymbol{\cdot}\bm{T}+\rho\,\mathbf{f}
	\label{eq:Cauchy}
\end{equation}

\noindent where $\rho$ is the density, $\mathbf{v}$ is the velocity vector, $\mathbf{f}$ is the body force per unit mass vector and $\mathbf{T}$ is the Cauchy stress tensor. Generally the Cauchy stress tensor is written as a function of the pressure p, and the extra-stress tensor $\bm{\sigma}$, as $\bm{T} = -p\bm{I}+\bm{\sigma}$, where $\bm{I}$ is the identity tensor. For incompressible fluids, this pressure is the part of the stress that is constitutively undetermined, while the extra-stress tensor is defined by a constitutive equation.

The closure equation involving the extra-stress tensor $\bm\sigma$ and the strain rate tensor $\bm{\dot\gamma}=\bm\nabla{\bm{v}} + \bm\nabla{\bm{v}^T}$ must be known in order to estimate the velocity and pressure fields. Newton's law of viscosity for incompressible fluid ($\bm\sigma=\mu\bm{\dot\gamma}$) is the simplest possible relation. The interplay between stress and strain rate differentiates fluid responses, highlighting the necessity of rheological constitutive equations in the formulation of momentum equations, with a notable emphasis on the importance of viscoelastic equations.

\subsection{Maxwell-Type Differential Constitutive Equations}
\label{Visco}

Most viscoelastic equations for fluids have been formulated extending Maxwell's original idea proposed in a landmark paper \cite{Maxwell1867} in 1867. We can broaden our comprehension of these Maxwell-type equations by examining an extended version of the Maxwell model \cite[~p.116]{Macosko1994}, given by

\begin{equation}
	\stackrel{\triangledown}{\bm{\sigma}}+\dfrac{\bm\sigma}{\lambda}+h(\bm\sigma) = G\bm{\dot\gamma}
	\label{eq:MT}
\end{equation}

In this work, we will explore equation (\ref{eq:MT}) with the four different models shown in table \ref{table:model}.

\begin{table}[ht]
\caption{Maxwell-type viscoelastic models. $\alpha$ is a mobility parameter, $\epsilon$ is the extensibility parameter and $\xi$ is affine/non-affine parameter.} 
\centering 
\begin{tabular}{c c c c c}
\hline\hline 
Model & $h(\bm\sigma)$ & $\xi$ & $\epsilon$ & $\alpha$\\ [0.5ex] 
\hline\\ 
UCM & 0 & - & - & - \\
Johnson and Segalman & $\dfrac{\xi}{2}(\bm\sigma\boldsymbol\cdot\bm{\dot\gamma}+\bm{\dot\gamma}\boldsymbol\cdot\bm\sigma)$ & (0,2] & - & - \\
Giesekus & $\dfrac{\alpha}{\lambda G} (\bm\sigma\boldsymbol{\cdot}\bm\sigma)$ & - & - &(0,1] \\[2ex] 
ePTT & $\dfrac{\bm\sigma}{\lambda}\left[\rm{exp}\left(\dfrac{\epsilon}{G}tr(\bm\sigma)\right)-\bm I\right]$	 & [0,2] & (0,1]&- \\[2ex] 
\hline
\end{tabular}
\label{table:model} 
\end{table}

In contrast to Newton's equation, the Maxwell-type constitutive equation (\ref{eq:MT}) is differential and involves other parameters besides viscosity. The relaxation time $\lambda$, which is an essential feature of elastic fluids. This parameter brings a memory characteristic, since it relates the time needed for the stresses to vanish ($\bm\sigma=0$) after the deformation rate is removed. 
The elastic modulus ($G$), governs the response to elastic deformation, specifically the material's resistance to deformation. Finally, $h(\bm\sigma)$ is a general function that assumes different forms depending on the constitutive equation. 

The symbol ($\stackrel{\triangledown}{}$)  in the equation (\ref{eq:MT}) is the upper convected time derivative operator introduced in Oldroyd's seminal work \cite{Oldroyd1950}, defined as

\begin{equation}
	\stackrel{\triangledown}{\bm{\sigma}} = \dfrac{\partial \bm\sigma}{\partial t} + \bm{v}\boldsymbol\cdot\bm\nabla \bm{\sigma} - (\bm\nabla \bm{v})^T\boldsymbol\cdot \bm{\sigma}- \bm{\sigma}\boldsymbol\cdot(\bm\nabla \bm{v})
	\label{eq:conv_op}
\end{equation}

If the function $h(\bm\sigma)$ in equation (\ref{eq:MT}) is given by $\frac{\xi}{2}(\bm\sigma\cdot\bm{\dot\gamma}+\bm{\dot\gamma}\cdot\bm\sigma)$, the Gordon-Schowalter derivative appears

\begin{equation}
	\stackrel{\square}{\bm{\sigma}} = \stackrel{\triangledown}{\bm{\sigma}}+\dfrac{\xi}{2}(\bm\sigma\boldsymbol\cdot\bm{\dot\gamma}+\bm{\dot\gamma}\boldsymbol\cdot\bm\sigma)
	\label{eq:gordon}
\end{equation}

being this operator a general convected time derivative that assumes, as particular cases, the upper convected time derivative ($\xi=0$), the co-rotational convected time derivative ($\xi=1$), and the lower convected time derivative ($\xi=2$). The Gordon-Schowalter operator is Euclidean invariant, i.e., frame indifferent, and essential condition required in elaborating constitutive equations from Continuum Mechanics \cite{Liu2002}. In addition, it enables shear thinning behavior \cite{Pivokonsky2015},\cite[~p.135 ]{Larson1988}. Otherwise, for $h(\bm\sigma)=\dfrac{\alpha}{\lambda G} (\bm\sigma\boldsymbol{\cdot}\bm\sigma)$ the Giesekus model is retrieved \cite{Giesekus1982} and $h(\bm\sigma)=\dfrac{\bm\sigma}{\lambda}\left[\rm{exp}\left(\dfrac{\epsilon}{G}tr(\bm\sigma)\right)-\bm I\right]$ the Exponential Phan-Thien-Tanner (ePTT) model \cite{Thien1978}. Both models, Giesekus and ePTT also exhibit shear-thinning. Setting $h(\bm\sigma)=0$ results in the established Upper Convected Maxwell (UCM) model, which has a constant shear viscosity.

Substitution of equation (\ref{eq:conv_op}) into equation (\ref{eq:MT}), yields:

\begin{equation}
	\dfrac{d\bm\sigma}{dt} = G\bm{\dot\gamma} -\dfrac{\bm\sigma}{\lambda}+  [(\bm\nabla \bm{v})^T\boldsymbol\cdot \bm{\sigma}+ \bm{\sigma}\boldsymbol\cdot\bm\nabla \bm{v}]-\bm v\boldsymbol\cdot\bm\nabla\bm\sigma -h(\bm\sigma)
	\label{eq:wt-final}
\end{equation}

Equation (\ref{eq:wt-final}) is a Maxwell-type general equation that can represent various viscoelastic models by adjusting the function $h(\bm\sigma)$, see table \ref{table:model}.

\section{Methodology}
\label{Metho}
\subsection{Universal Differential Equation}

Formulating a constitutive equation is a complex task that requires a deep understanding of how fluids respond to various stress conditions. New approaches have been explored to derive these equations, particularly techniques that utilize data-driven techniques that adhere to the underlying physics of the phenomena. The Universal Differential Equation \cite{Rackauckas2021} is a type of physics-informed data-driven method based on a seminal paper of \citet{Chen2018}, which illustrated that some types of neural networks (such as Residual and Recurrent Network) can be interpreted as discretized ordinary differential equations, like with the explicit Euler method, transforming them into initial value problem of Neural Ordinary Differential Equations (NODEs), of the form

\begin{equation}
	\dfrac{du}{dt}=\mathscr{N}_{\theta}(u,t),
	\label{eq:NODE}
\end{equation}
i.e. described by a neural network $\mathscr{N}_{\theta}$, in which $\theta$ represents the weights. By way of example, a neural network with two hidden layers is represented by

\begin{equation}
	 \mathscr{N}_{\theta}(u,t) = W_3\phi_2\left(W_2\phi_1\left(W_1\begin{bmatrix} u \\ t \end{bmatrix} + b_1\right) + b_2\right) + b_3,
	\label{eq:neural-network}
\end{equation}
where $\theta = (W_1, W_2, W_3, b_1, b_2, b_3)$, $W_i$ are matrices of weights, $b_i$ are vectors of biases, and $(\phi_1, \phi_2)$ are activation functions. Nonetheless, the model produced by NODEs does not assimilate pre-existing knowledge based on known mechanics from first principles.

To tackle the limitations of NODEs, Rackauckas et al.\cite{Rackauckas2021} introduced the Universal Differential Equation (UDE) methodology. This approach combines mechanistic modeling with a universal approximator, such as a neural network, to facilitate flexible data-driven model enhancements. Usually, a neural network is used to adjust a differential equation, which is the mechanistic model that helps to recover the missing terms of the equation. The described methodology has a hybrid character since it combines prior physical knowledge in the form of a differential equation with a data-driven machine learning algorithm. It can be expressed by

\begin{equation}
	\dfrac{d\bm\sigma}{dt}=f(\bm\sigma,\bm{\dot\gamma}, t,\mathscr{N}_{\theta}(\bm\sigma,\bm{\dot\gamma},t)).
	\label{eq:UDE}
\end{equation}

In this case study, $f$ represents a particular function that describes the behavior of the fluid based on known information, namely, a viscoelastic constitutive equation. It is defined by equation (\ref{eq:wt-final}) where the function $h(\bm\sigma)$ is yet to be determined.

The neural network, denoted as $\mathscr{N}$ with parameter $\theta$, is a function of $\bm\sigma$ and $\bm{\dot\gamma}$ at each time $t$, along with other relevant parameters.

In this paper, we examine the application of equation (\ref{eq:wt-final}), considering that certain terms are unknown, in order to explore the process of reconstructing the underlying models in Table \ref{table:model}. The unknown can term differ depending on the model chosen. Given that rheometric experiments are generally carried out in small gaps, we investigate homogeneous flow $(\bm\nabla\bm\sigma=0)$ in all scenarios. In this work the use of dimensionless form of equation (\ref{eq:wt-final}), and function $h(\bm\sigma)$ provided in table \ref{table:model}, considering $t^*=t/\lambda$. $\bm\sigma^{*}=\bm\sigma/G$; $\bm {\bm{\dot\gamma}}^*=\lambda\bm{\dot\gamma}$; $(\bm \nabla \bm v)^{*} = \lambda\bm \nabla \bm v$. With such approach, the solution is not influenced by the values of $G$ and $\lambda$.

The differential equation for the Johnson-Segalman and Giesekus models can be summarized in the equation

\begin{align}
    \centering    
	&\dfrac{d\bm\sigma^*}{dt^*} = {\bm{\dot\gamma}^*} -\bm{\sigma^*}  + {(\bm\nabla \bm{v})^{*}}^T\boldsymbol\cdot \bm{\sigma^*}+\bm{\sigma^*}\boldsymbol\cdot(\bm\nabla\bm{v})^{*}-\mathscr{N}(\bm\sigma^*,{\bm{\dot\gamma}})
	\label{eq:universal-adim-JS-GS}
\end{align}

In this case, the neural network will be responsible for the recovery of the terms, $\mathscr{N}(\bm\sigma^*,{\bm{\dot\gamma}}^*)\equiv\dfrac{\xi}{2}(\bm\sigma^*\boldsymbol\cdot{\bm{\dot\gamma}}^*+{\bm{\dot\gamma}}\boldsymbol\cdot\bm\sigma^*)$ for Johnson-Segalman, and $\mathscr{N}(\bm\sigma^*,{\bm{\dot\gamma}}^*)\equiv\alpha(\bm\sigma^*\boldsymbol\cdot\bm\sigma^*)$ for Giesekus.

For the UCM and ePTT models, the equation employed is given by

\begin{align}
    \centering   
	&\dfrac{d\bm\sigma^*}{dt^*} = {\bm{\dot\gamma}}^* + (\bm\nabla {\bm{v})^*}^T\boldsymbol\cdot \bm{\sigma^*}+\bm{\sigma^*}\boldsymbol\cdot(\bm\nabla\bm{v})^*-\mathscr{N}(\bm\sigma^*,{\bm{\dot\gamma}}^*)
	\label{eq:universal-adim-UCM-PTT}
\end{align}
where $\mathscr{N}(\bm\sigma^*,{\bm{\dot\gamma}}^*)\equiv\bm\sigma^*$ for UCM and $\mathscr{N}(\bm\sigma^*,{\bm{\dot\gamma}}^*)\equiv\bm\sigma^*\rm{exp}\left[\epsilon\, tr(\bm\sigma^*)\right]$ for ePTT. 

As the extra-stress tensor is symmetric ($\bm\sigma^*={\bm\sigma^*}^T$), equations (\ref{eq:universal-adim-JS-GS}) and (\ref{eq:universal-adim-UCM-PTT}) give rise to a system of six differential equations for six unknowns $\sigma^*_{11}$, $\sigma^*_{22}$, $\sigma^*_{33}$, $\sigma^*_{12}$, $\sigma^*_{13}$, and $\sigma^*_{23}$.

\subsection{Neural Network Architecture}

The differential equation employed as \textit{prior} knowledge is frame-indifferent. The neural network must preserve this property to maintain the invariance specified by the differential equation. To accomplish this, we employ the network proposed by \citet{Ling2016}, which \citet{Lennon2023} also applied. Firstly, a scalar coefficient $g$ is obtained from a multilayer perceptron (MLP) neural network's output, which is then organized as a Tensor Basis Neural Network (TBNN) applying the Hadamard product $\odot$ by a tensor ${\bm{T}^{*}}$. Figure \ref{fig:TBNN} shows the neural network architecture.

\begin{figure}[H]
	\centering
	\includegraphics[width=1\linewidth]{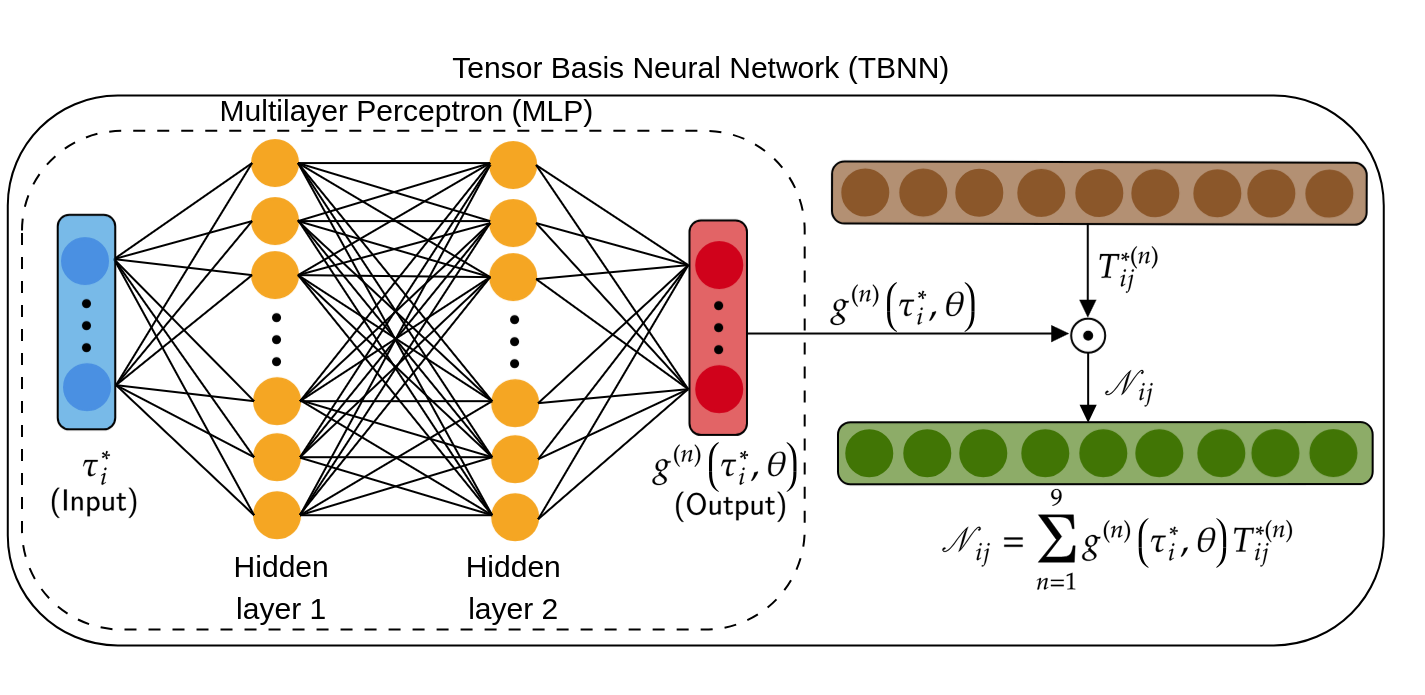}
	\caption{Tensor basis neural network architecture (TBNN).  The invariants ($\tau_i^*$) of the basis tensor ${T_{ij}^*}^{(n)}$ are the inputs. The output of the multilayer perceptron is a scalar coefficient $g^{(n)}(\tau_i^*;\theta)$. $\mathscr{N}_{ij}$ is the final result produced by the Hadamard product (element-wise) $\odot$ between $g^{(n)}(\tau_i^*;\theta)$ and ${T_{ij}^*}^{(n)}$. The MLP model is composed of four layers: an input and output layer with nine neurons each and two hidden layers with 32 neurons each that utilize the tanh activation function.}
	\label{fig:TBNN}	
\end{figure} 

The Hadamard product (element-wise) for each component of $\mathscr{N}_{ij}(\bm\sigma^*,{\bm{\dot\gamma}}^*)$ is defined by

\begin{equation}
	\mathscr{N}_{ij}(\bm\sigma^*,{\bm{\dot\gamma}}^*) = g\odot\bm {T^*}=\sum_{n=1}^{8}g^{(n)}(\{\tau_i^*\}_{i=1,2,3...,8};\theta){\mathbf{T}_{ij}^{*}}^{(n)}
	\label{eq:poly_exp}
\end{equation}

The final neural network's output, the components of $\mathscr{N}_{ij}(\bm\sigma^*,{\bm{\dot\gamma}}^*)$, is defined by a set of $n$ finite tensor polynomials, where $g^{(n)}$ represents a scalar coefficient that depends on the neural network parameters and eight invariants ${\tau_i}^*$ of ${\mathbf{T}_{ij}^{*}}^{(n)}$, which represent basis tensors that depend on two symmetric tensors, $\bm\sigma^*$ and ${\bm{\dot\gamma}}^*$, thereby ensuring the Euclidean invariance. In contrast to \citet{Lennon2023}, we use Smith's \cite{Smith1971} representation theorem to set the number of tensors at eight. We performed a preliminary study and verified that the extra terms considered by \citet{Lennon2023} were unnecessary. Equations (\ref{eq:invariants}) and (\ref{eq:tensors}) show the invariants $\tau_i^*$ and tensors ${\mathbf{T}_{ij}^{*}}^{(n)}$. 

\begin{equation}
	\hspace{-2cm}\tau_i^*=	
	\begin{cases}	
		\begin{aligned}
			\tau_1^* &= tr(\bm{\sigma^*})\\
			\tau_2^* &= tr(\bm{\sigma^*}\boldsymbol\cdot\bm{\sigma^*})\\
			\tau_3^* &= tr({\bm{\dot\gamma}}^*\boldsymbol\cdot{\bm{\dot\gamma}}^*)\\
			\tau_4^* &= tr(\bm{\sigma^*}\boldsymbol\cdot\bm{\sigma^*}\boldsymbol\cdot\bm{\sigma^*})\\
			\tau_5^* &=tr({\bm{\dot\gamma}}^*\boldsymbol\cdot{\bm{\dot\gamma}}^*\boldsymbol\cdot{\bm{\dot\gamma}}^*)\\		
			\tau_6^* &= tr(\bm{\sigma^*}\boldsymbol\cdot\bm{\sigma^*}\boldsymbol\cdot{\bm{\dot\gamma}}^*)\\
			\tau_7^* &= tr(\bm{\sigma}^*\boldsymbol\cdot{\bm{\dot\gamma}}^*\boldsymbol\cdot{\bm{\dot\gamma}}^*)\\
			\tau_8^* &= tr(\bm{\sigma^*}\boldsymbol\cdot{\bm{\dot\gamma}}^*)\\\\		
		\end{aligned}
	\end{cases}	
	\label{eq:invariants}
\end{equation}

\begin{equation}
	{\mathbf{T}_{ij}^{*}}^{(n)}=	
	\begin{cases}	
		\begin{aligned}
		{\mathbf{T}_{ij}^{*}}^{(1)}&= \mathbf{I}\\
		{\mathbf{T}_{ij}^{*}}^{(2)}&= \bm{\sigma^*}\\
		{\mathbf{T}_{ij}^{*}}^{(3)} &= {\bm{\dot\gamma}}^*\\
		{\mathbf{T}_{ij}^{*}}^{(4)} &= \bm{\sigma^*}\boldsymbol\cdot\bm{\sigma^*}\\
		{\mathbf{T}_{ij}^{*}}^{(5)}&= {\bm{\dot\gamma}}^*\boldsymbol\cdot{\bm{\dot\gamma}}^*\\		
		{\mathbf{T}_{ij}^{*}}^{(6)} &= \bm{\sigma^*}\boldsymbol\cdot{\bm{\dot\gamma}}^*+{\bm{\dot\gamma}}^*\boldsymbol\cdot\bm{\sigma^*}\\
		{\mathbf{T}_{ij}^{*}}^{(7)} &= \bm{\sigma^*}\boldsymbol\cdot\bm{\sigma^*}\boldsymbol\cdot{\bm{\dot\gamma}}^*+{\bm{\dot\gamma}}^*\boldsymbol\cdot\bm{\sigma^*}\boldsymbol\cdot\bm{\sigma^*}\\
		{\mathbf{T}_{ij}^{*}}^{(8)}&= \bm{\sigma^*}\boldsymbol\cdot{\bm{\dot\gamma}}^*\boldsymbol\cdot{\bm{\dot\gamma}}^*+{\bm{\dot\gamma}}^*\boldsymbol\cdot{\bm{\dot\gamma}}^*\cdot\bm{\sigma^*}\\			
		\end{aligned}
	\end{cases}
	\label{eq:tensors}	
\end{equation}

where $\bm I$ represents the identity tensor and $tr({\bm{\dot\gamma}}^*)$ is not considered since it vanishes in incompressible fluids.
Once the MLP network has determined the coefficients $g^{(n)}$, we construct all the components of the tensor $\mathscr{N}_{ij}$ through the polynomial expansion represented in equation (\ref{eq:poly_exp}). Subsequently, equation (\ref{eq:universal-adim-JS-GS}) or (\ref{eq:universal-adim-UCM-PTT}) is solved using an integration method at each time step to obtain the stress field. The MLP architecture consists of four layers, including an input and output layer containing nine neurons each, along with two hidden layers comprising 32 neurons each that make use of the hyperbolic tangent (tanh) as an activation function.

\subsection{Dataset and Training Procedure}
The model will be trained through a continuous process of incremental data addition, which is a kind of incremental learning \cite{Ven2022}, meaning that for every iteration $k$, a new time series will be included in the training set. Consequently, the initial series will be utilized for training until $k$ iterations have been completed, at which point a second series will be added. Training continues for another $k$ iterations with both series and so forth until training is finished with all the time series. This training procedure is the same presented by \citeauthor{Lennon2023}\cite{Lennon2023}.

In this study, the dataset was generated considering an oscillatory shear flow $v_1=\omega \gamma_0 \rm{cos}(\omega t)x_2\hat{e}_1$, where $x_1$ is the velocity direction and $\hat{e}_1$ is the corresponding unit vector, $x_2$ is the coordinate perpendicular to the wall, where changes in velocity occur and $\gamma_0$ is the strain amplitude. The input velocity gradient in dimensionless form is written as $\bm \nabla \bm v^*_{21} = \dot\gamma^*(t^*)=$ Wi\,\rm{cos}\,(De\,$t^*)$, with Deborah number values De $\equiv \lambda\omega \equiv$[0.25,0.5,0.75,1] and Weissenberg number Wi $\equiv \lambda\gamma_0\omega \equiv$[1,2], generating eight different combinations of input shear rate time series ($t^*$,  ${\dot\gamma}_{12}^*$). Since the values of the Weissenberg number are not small, these tests are in the realm of large amplitude oscillatory shear (LAOS).  The dataset was then generated from the numerical solution of equations (\ref{eq:universal-adim-JS-GS}) and (\ref{eq:universal-adim-UCM-PTT}) by Tsitouras 5/4 Runge-Kutta (Tsit5) method \cite{Tsitouras2011}, which includes an adaptive time step. Considering each of the four models evaluated in the present work, a total of 32 shear stress time series ($t^*$, $\sigma_{12\,(data)}^*$) were created for training purposes. 
Equations (\ref{eq:universal-adim-JS-GS}) and (\ref{eq:universal-adim-UCM-PTT}) were implemented in Julia language in the \texttt{DifferentialEquations.jl} library \cite{Rackauckas2017}. The initial condition for all experiments was zero stress field $\bm\sigma^*=0$ and $t^* \in [0,20]$.
It should be emphasized that the training data exclusively consists of the time series of the shear stress component ($\sigma^*_{12}$), with no inclusion of normal components ($\sigma^*_{11},\sigma^*_{22},\sigma^*_{33}$) data or shear components data associated with the neutral direction $x_3$ ($\sigma^*_{13}$,$\sigma^*_{23}$) at any given time. A total of 7200 iterations were conducted to conclude the optimization process, with each incremental series encompassing 900 iterations.

\subsubsection{Differentiable Physics and Optimization Process}

Working with neural networks eventually leads to an optimization problem. Minimizing the loss function $\mathcal{L}$ is crucial when updating the neural network weights.
The loss function applied in this work employs $L^1$ regularization, i.e.,
\begin{equation}
	\mathcal{L}(\hat\sigma^*_{12}(t_i;\theta),\theta) = \sum_{E_j\in\{E\}}\frac{1}{N}\sum_{i=1}^N\lvert\lvert\sigma^{*}_{12\,(data)}(t_i;\theta)-\hat{\sigma}^{*}(t_i;\theta)\lvert\lvert_2^2 + \kappa\lvert\lvert\theta\lvert\lvert_1
	\label{eq:loss-function}
 \end{equation}
 
 \noindent where $\hat\sigma^*_{12}(t)=\texttt{EDOSolver}(\hat\sigma^{*}_{12}(t_0),\mathscr{N}_{\theta},t_0,t_f,\theta)$ represents the value predicted by the model at the $i$-th corresponding point of the time series, obtained by means of a numerical integration method from an initial condition $t_0$ and time of analysis $t_f$. The set of experiments is denoted as $\{E\}$, with $E_j$ representing the $j$-th experiment out of a total of eight. The parameter $\kappa$ is a regularization coefficient for the weights $\theta$, adopted in this work as $10^{-2}$, and $N$ is the total number of points in the set $\{E\}$. 

 During training, there occurs a process of minimization of the loss function  $\mathcal{L}(\hat\sigma_{12}(t_i;\theta)$ defined in terms of the current solution  $\hat\sigma^*_{12}(t,\theta)$ of the hybrid differential equation and the current choice of the parameters $\theta$. 
 
If a gradient-based method (like ADAM \cite{Kingma2014}) is employed to minimize the loss function, it is essential to use a robust procedure to compute the gradient of the loss function. 

The computational complexity of computing gradients directly backpropagating through operations of the EDO solver increases significantly for sophisticated solvers. The continuous adjoint sensitivity analysis (CASA) technique has been used to circumvent this limitation instead and will be applied in this study \cite{Sapienza2024}. This technique is divided into three steps: 
 \begin{enumerate}
     \item Solve the differential equation (\ref{eq:UDE}) forward in time;
     \item Consider the continuous version of the loss function 
     \[L(\sigma;\theta)=\int_{t_0}^{t_f}\ell(\sigma(t;\theta),\theta)dt\]
     being $\ell$  a function to measure the significance of the error component at each time $t$ within the interval [$t_0$, $t_1$]. Herein, $\ell$ is defined according to general terms in the summation of the equation (\ref{eq:loss-function})\[\ell=\frac{1}{N}\lvert\lvert\sigma^{*}_{12\,(data)}(t;\theta)-\hat{\sigma}^{*}(t;\theta)\lvert\lvert_2^2 + \kappa\lvert\lvert\theta\lvert\lvert_1\]   
     Then, solve backward in time from $t_f$ to $t_0$ an adjoint differential equation
     \begin{equation}
     \dfrac{d\lambda}{dt}=-\left(\dfrac{\partial f}{\partial \sigma}\right)^T\lambda-\left(\dfrac{\partial \ell}{\partial \sigma}\right)^T,\,\lambda(t_f)=0
     \label{eq:adjoint}
     \end{equation}
     \noindent where $\lambda(t) \in \mathbb{R}^n$ is the Lagrange multiplier of the continuous
     constraint defined by the differential equation (\ref{eq:UDE}); 
     \item Compute the gradient according to
      \begin{equation}
     \dfrac{dL}{d\theta}=\lambda^T(t_0)s(t_0)+\int_{t_0}^{t_f}\left(\dfrac{\partial \ell}{\partial \theta}+\lambda^T\dfrac{\partial f}{\partial \theta}\right)dt
     \label{eq:full_gradient}
     \end{equation}
     where $s=\frac{\partial\sigma}{\partial\theta}$ is the sensitivity.
 \end{enumerate}

 Understanding the numerical solution of $\sigma(t)$ at each time and how to handle the vector-Jacobian products (VJPs), defined as 
 $$\int_{t_0}^{t_f}\frac{\partial \ell}{\partial\sigma}s(t)= \lambda^T(t_0)s(t_0)+\int_{t_0}^{t_f}\lambda^T\frac{\partial f}{\partial \theta}dt,$$ 
 is essential for solving the adjoint equation (\ref{eq:adjoint}) and (\ref{eq:full_gradient}) respectively. An interpolation method is used to store in memory the intermediate states of $\sigma(t)$ during the forward solution of equation (\ref{eq:UDE}) and thus being applied for the reverse solver vector-Jacobian products of equation (\ref{eq:full_gradient}). The library \texttt{SciMLSensitivity.jl} \cite{Rackauckas2021} available in Julia allows us to perform these calculations by calling the \texttt{InterpolatingAdjoint} and \texttt{ReverseDiffVJP} functions, respectively.

Upon completion of steps 1 to 3, the weights optimization procedure of the neural networks is carried out utilizing the ADAM's algorithm \cite{Kingma2014} at a specified learning rate of $10^{-3}$ and exponential decay rates for the moment estimates of $(\beta_1,\beta_2)=(0.9,0.999)$. Optimization was carried out with \texttt{Optimization.jl} package in Julia \cite{Kumar2023}. The methodology is summarized in the algorithm described in the flowchart of Figure \ref{fig:flowchart}.

\begin{figure}[H]
	\centering
	\includegraphics[width=1.0\linewidth]{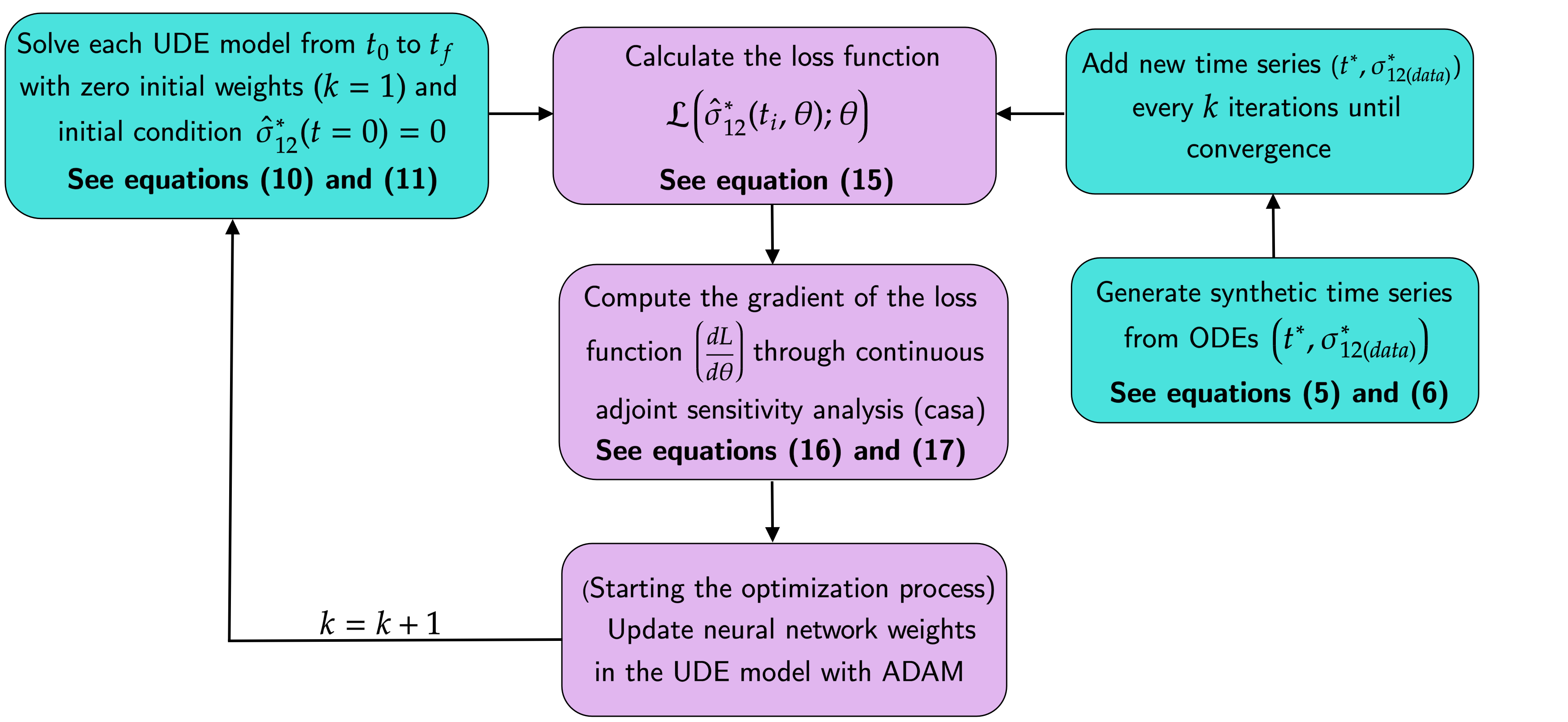}
	\caption{Flowchart of the algorithm to solve UDE. At $k=900$ iterations, a new time series is added in the optimization process until all eight-time series have been trained $(
k=7000)$.}
	\label{fig:flowchart}	
\end{figure}

\section{Results}
\subsection{Viscoelastic modeling}

Three experiments were carried out to test the extrapolation of the UDE model, namely, two oscillatory experiments with an strain rate input $\dot\gamma^*(t^*)=$ Wi\,cos\,(De\,$t^*$), taking the pairs (Wi,De)=[(3,1.5);(2,1)] and one experiment taking constant strain rate of $\dot\gamma^{*} = 2$. Figure \ref{fig:Loss} shows the training results for each model. The observed peaks correspond to the insertion of a new series during the optimization process. Convergence occurs around iteration 5000. 

\begin{figure}[H]
	\centering
	\includegraphics[width=0.7\linewidth]{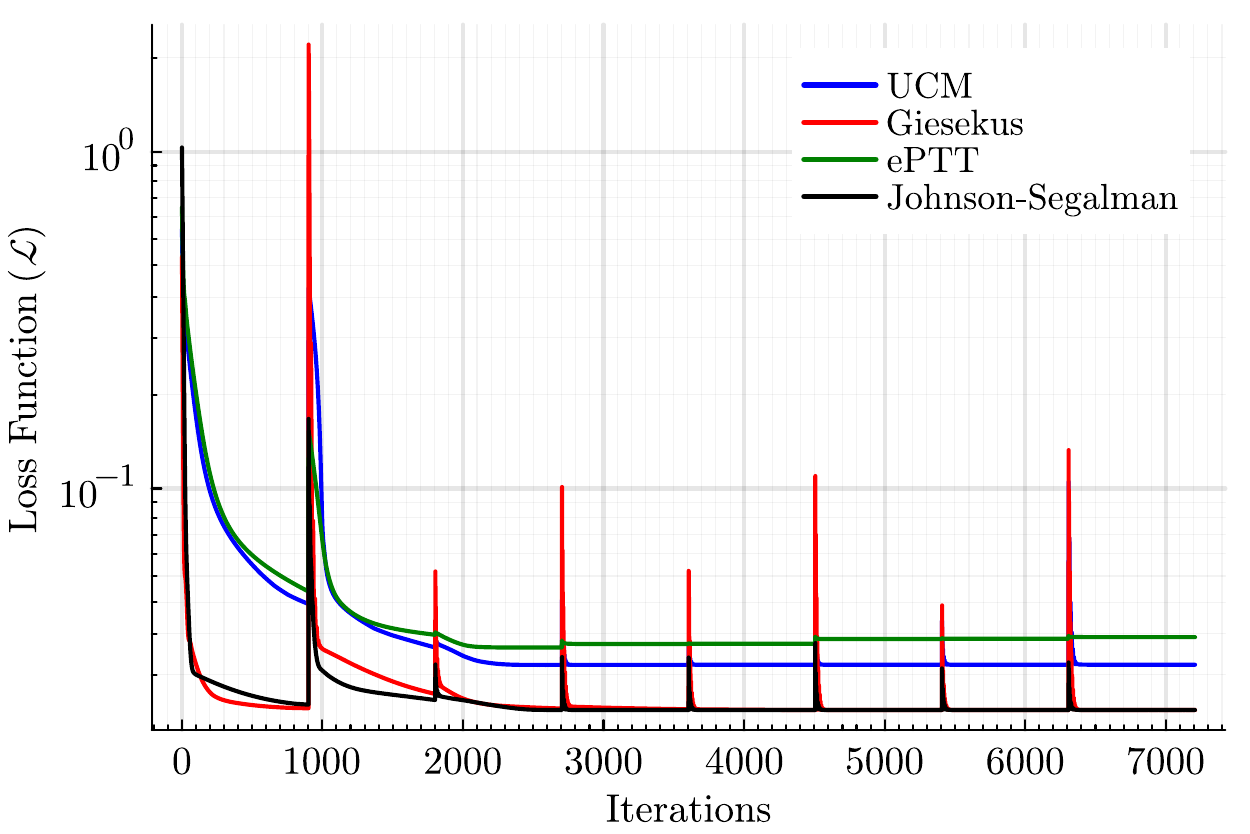}
	\caption{Loss function for each model used as \textit{a priori} information in the UDE.}
	\label{fig:Loss}	
\end{figure} 

\begin{figure}[htb!]
	\centering
	\includegraphics[width=0.9\linewidth]{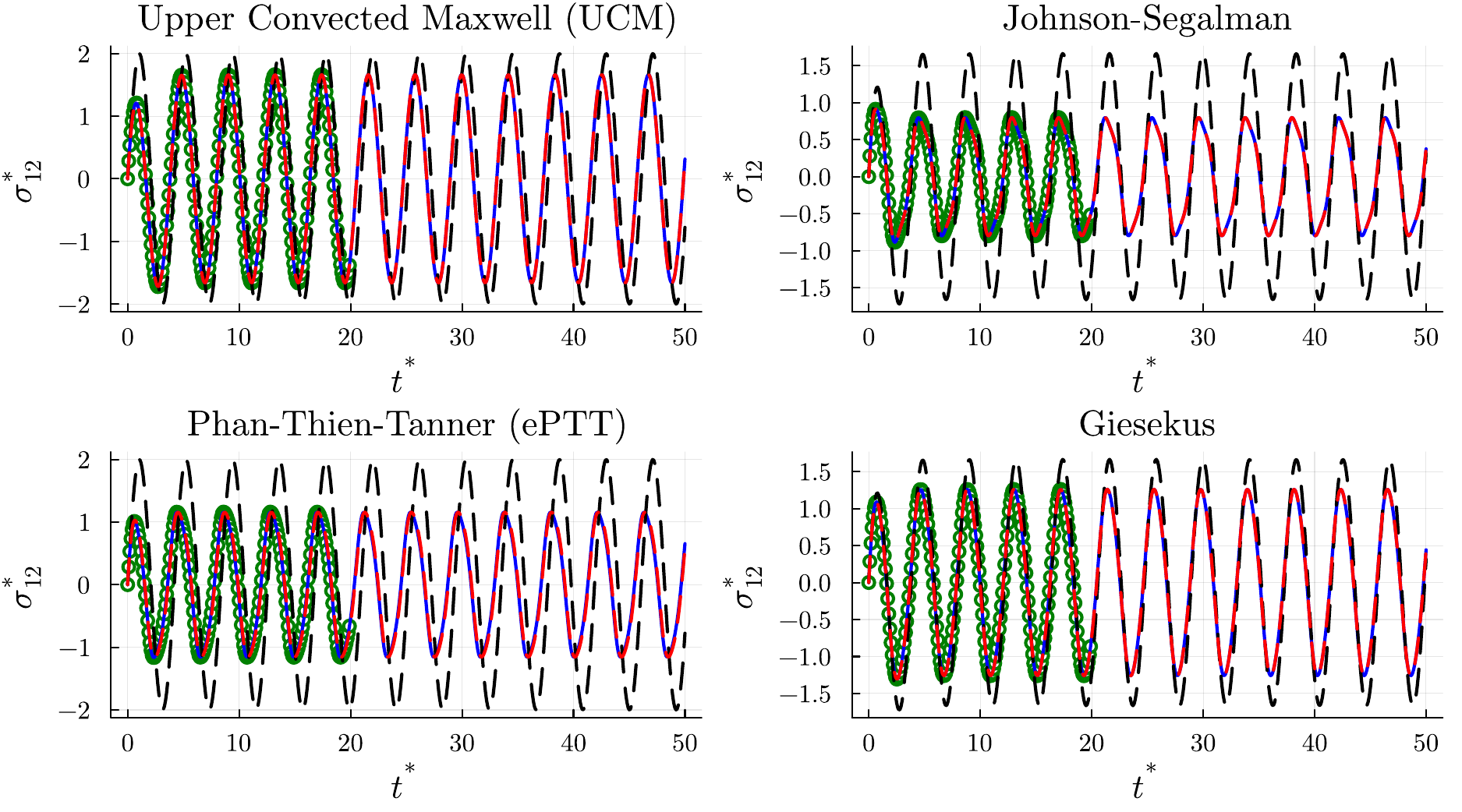}
	\caption{Evaluation for the extrapolation of UDE models for shear stress ($\sigma^{*}_{12},t^*$) with input $\dot\gamma^*(t^*)=$ 3\,cos\,(1.5\,$t^*$). The points highlighted in green correspond to training points, $t^*\in[0,20]$. The blue curve illustrates the numerical solution of the differential equation, referred to as the "ground truth" solution. The black curve depicts the UDE model pre-training, while the red curve illustrates the UDE model post-training. The following parameters were used $\xi=0.4$ (Johnson-Segalman), $\alpha=0.2$ (Giesekus) and ($\xi$, $\epsilon$)=(0,0.4) for (ePTT).}
	\label{fig:plot_1}
\end{figure}

Figure \ref{fig:plot_1} shows the results of the extrapolation tests of the UDE models with input $\dot\gamma^*(t^*)=$ 3\,cos\,(1.5\,$t^*$) and output  $\sigma^{*}_{12}$ with the choice of parameters: $\xi=0.4$ (Johnson-Segalman), $\alpha=0.2$ (Giesekus) and ($\xi$, $\epsilon$)=(0,0.4) for (ePTT). In this case, the model predictions aimed for times longer than the ones employed in the training ($t^{*}>20$) stage.  The blue curves illustrate the numerical solution of the differential equation, referred to as the "ground truth" solution. The black curves depict the UDE model pre-training, i.e., $\mathscr{N}_{ij}(\bm\sigma^*,{\bm{\dot\gamma}}^*)=0$, while the red curves represent the UDE model post-training.

The coefficients  $g^{(n)}(\tau^{*},\theta)$ within the network, as delineated by equation \ref{eq:poly_exp}, demonstrated values on the order of $10^{-7}$ to $10^{-4}$, apart from the coefficients linked to $T^{(2)}_{ij}$ (coefficient $g^{(2)}$) for the UCM and ePTT models,  $T^{(4)}_{ij}$ (coefficient $g^{(4)}$), $T^{(6)}_{ij}$ (coefficient  $g^{(6)}$) for Giesekus and Johnson-Segalman models, respectively. The values of the coefficients $g^{(n)}$ found are equivalent to 0.99 (proper 1) for UCM, 0.199 (proper 0.2) for Johnson-Segalman, and 0.199 (proper 0.2) for Giesekus. Consequently, the tensors $\bm\sigma^*$, $\frac{\xi}{2}$$(\bm\sigma^{*}\boldsymbol\cdot{\bm{\dot\gamma}}^*+{\bm{\dot\gamma}}^*\boldsymbol\cdot\bm\sigma^{*})$, $\alpha(\bm\sigma^{*}\boldsymbol\cdot\bm\sigma^{*})$ for the aforementioned models could be retrieved. An analysis of the coefficients of the ePTT model shall be undertaken at a later point in the text. 

Regardless of the model used to generate the synthetic data, the universal differential equation could capture the behavior from the training dataset, whether in time, amplitude, or frequency. The UDE model was exclusively trained using shear stress ($\sigma^{*}_{12}$) data collected from oscillatory experiments. In the following stage, the model will assessed by predicting normal stress differences obtained from information not used during the model's training. 

Figure \ref{fig:plot_2} illustrates the investigation of the first normal stress difference in shear, N$_1^* \equiv \sigma^*_{11} - \sigma^*_{22}$. Despite the model not being trained on the normal stress data, it delivered accurate predictions, except for the ePTT model, which manifested a slight amplitude discrepancy between the exact ("ground truth") solution (blue curve) and the estimation (red curve).
The determination of the second normal stress difference in shear, N$^{*}_2$, from experimental measurements is notoriously difficult \cite{Maklad2021}. The depiction in Figure \ref{fig:plot_3} illustrates the changes in N$^{*}_2$ for each model. The predictions for the Johnson-Segalman and Giesekus models exhibited excellent agreement with the original corresponding solutions for N$^{*}_2$. Even though the predicted UCM  presented non-zero values, in contrast to the original models, they are much smaller (by three and two orders of magnitude) compared to N$^{*}_1$, indicating that the ratio N$^{*}_2/$N$^{*}_1$ is nearly zero. The ePTT model exhibited  non-zero values and marginally positive values in the initial stages, a topic that will be elaborated on further in the text.

\begin{figure}[H]
	\centering
	\includegraphics[width=0.9\linewidth]{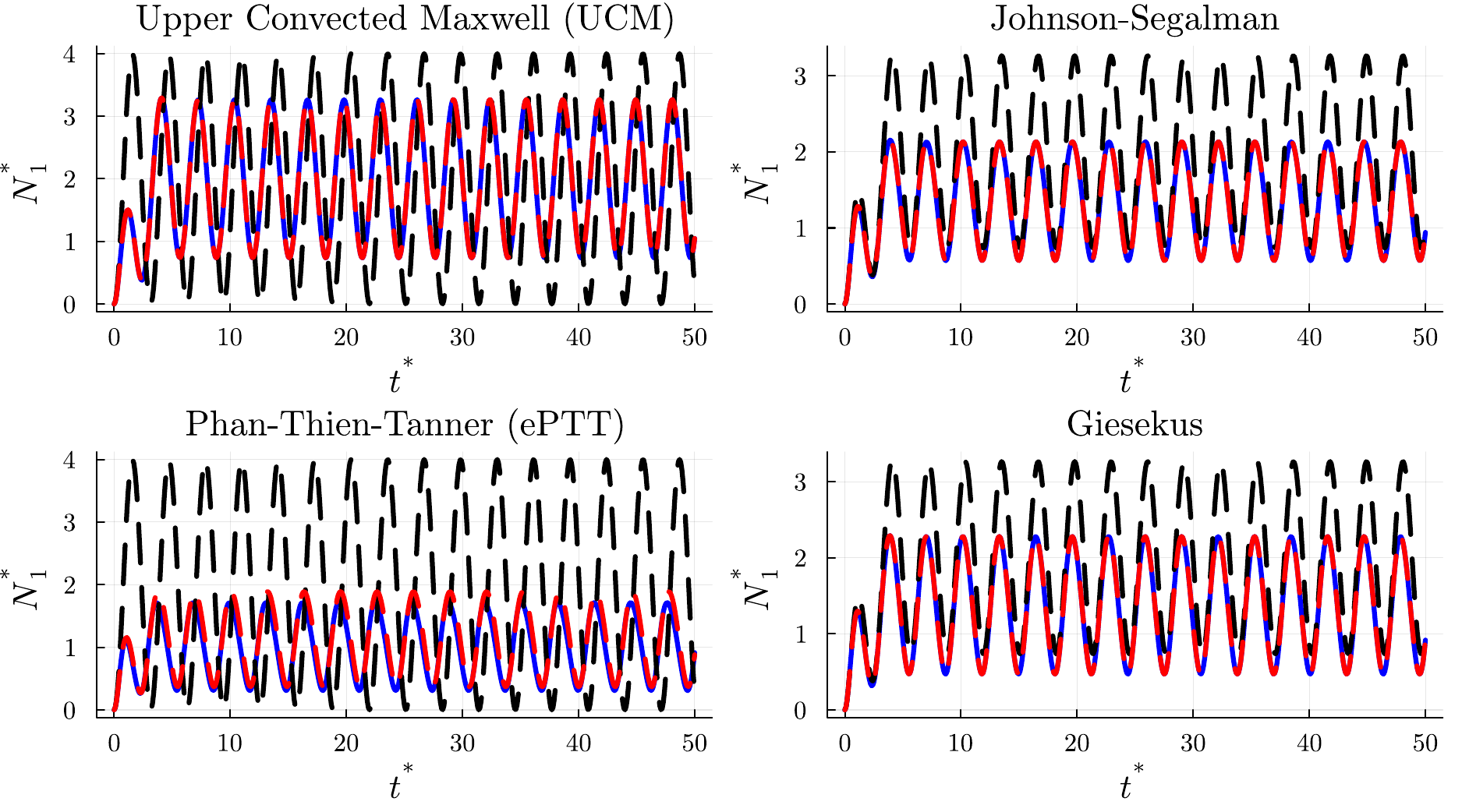}
	\caption{Evaluation for the extrapolation of UDE models for first normal stress difference in shear (N$_1^* = \sigma^*_{11} - \sigma^*_{22}$)(N$_1^*,t^*$) with input $\dot\gamma^*(t^*)=$ 2\,cos\,$t^*$. The blue curve illustrates the numerical solution of the differential equation, referred to as the "ground truth" solution. The black curve depicts the UDE model pre-training, while the red curve illustrates the UDE model post-training. The following parameters were used $\xi=0.4$ (Johnson-Segalman), $\alpha=0.2$ (Giesekus) and ($\xi$, $\epsilon$)=(0,0.4) for (ePTT).}
	\label{fig:plot_2}
\end{figure}
\begin{figure}[H]
	\centering
	\includegraphics[width=0.9\linewidth]{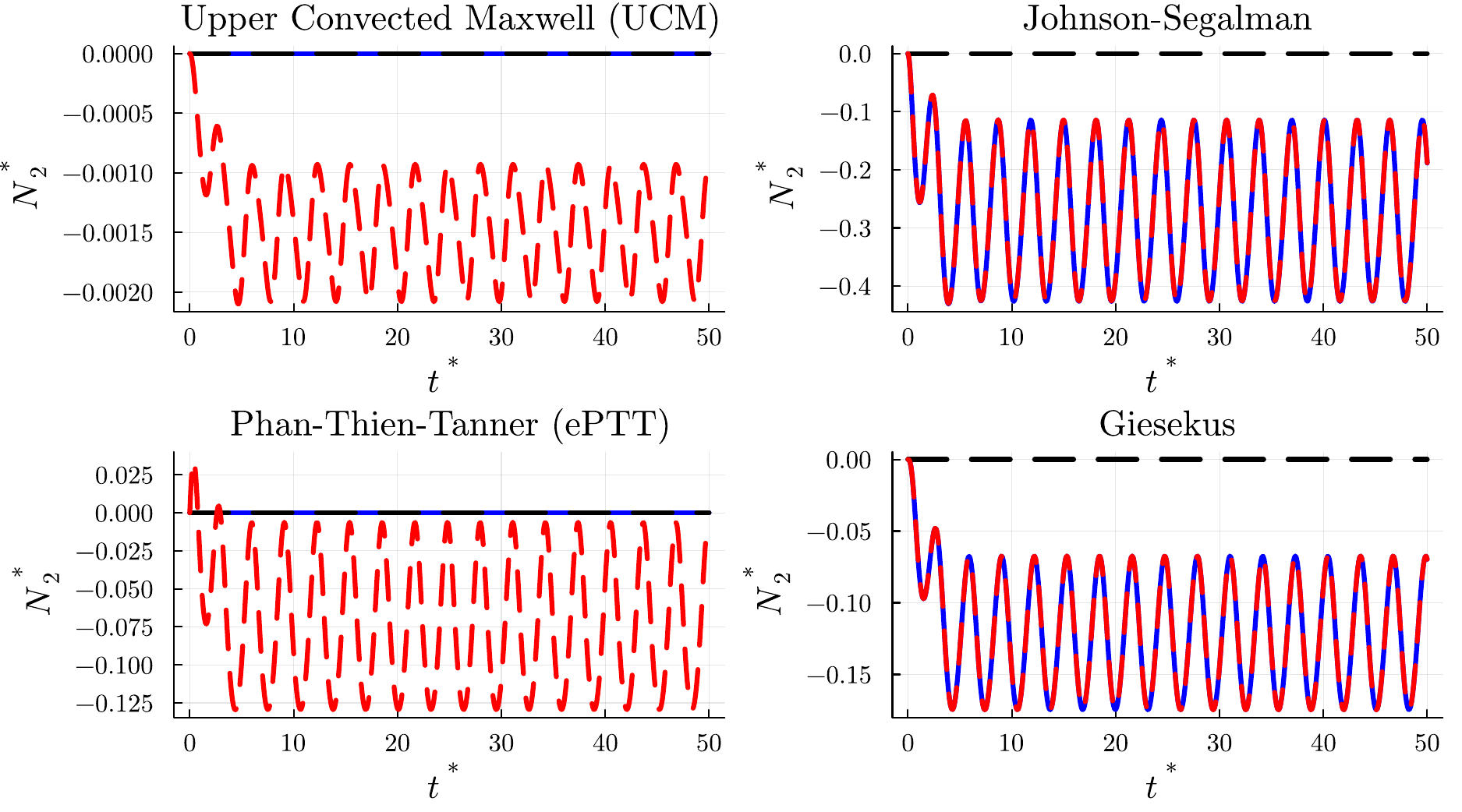}
	\caption{Evaluation for the extrapolation of UDE models for second normal stress difference (N$_2^* = \sigma^*_{22} - \sigma^*_{33}$)(N$_2^*,t^*$) with input $\dot\gamma^*(t^*)=$ 2\,cos\,$t^*$. The blue curve illustrates the numerical solution of the differential equation, referred to as the "ground truth" solution. The black curve depicts the UDE model pre-training, while the red curve illustrates the UDE model post-training. The following parameters were used $\xi=0.4$ (Johnson-Segalman), $\alpha=0.2$ (Giesekus) and ($\xi$, $\epsilon$)=(0,0.4) for (ePTT).}
	\label{fig:plot_3}
\end{figure}

The viscous Lissajous curves display the preceding results in phase space as depicted in Figures \ref{fig:plot_5}, \ref{fig:plot_6}, and \ref{fig:plot_7}. The analysis of these cyclic curves in the $\sigma^*$ $\times$ ${\dot{\gamma}}^*$ domain helps to identify viscoelastic behavior and non-linearities \cite{Collyer1998}. In Fig.~\ref{fig:plot_5} the black elliptical lines are associated with the baseline cases (pre-training), which for Giesekus and Johnson-Segalman are exactly the UCM case (see Eq.~(\ref{eq:universal-adim-JS-GS})), while for the UCM and ePTT models we notice a purely elastic behavior (see Eq.~\ref{eq:universal-adim-UCM-PTT}) where the shear stress is out-of-phase with respect to the shear rate. It is worth noticing a transient stage that begins in time $t^*=0$ where $\dot{\gamma}^*=3$ and $\sigma^*_{12}=0$, where the trajectory in the Lissajous curves have not reached a closed cycle. Figure \ref{fig:plot_5} shows that all predicted models (red curve) are in good agreement with the ground truth solution (blue curve). 

\begin{figure}[H]
	\centering
	\includegraphics[width=0.9\linewidth]{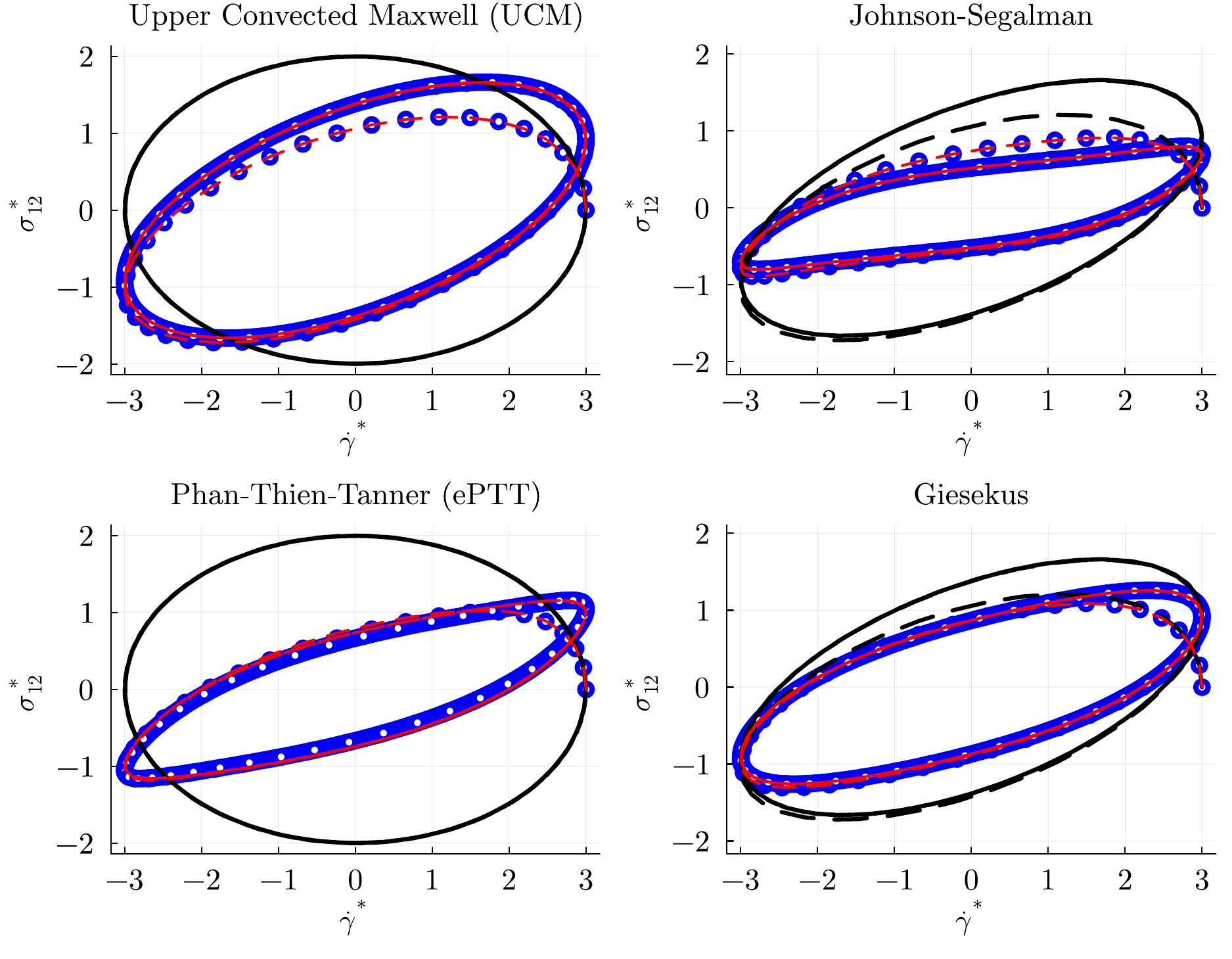}
	\caption{Lissajous-Bowditch curves for the extrapolation of UDE models for shear stress ($\sigma^{*}_{12},t^*$) with input $\dot\gamma^*(t^*)=$ 3\,cos\,(1.5\,$t^*$). The blue curve illustrates the numerical solution of the differential equation, referred to as the ground truth solution. The black curve depicts the UDE model pre-training, while the red curve illustrates the UDE model post-training. The following parameters were used $\xi=0.4$ (Johnson-Segalman), $\alpha=0.2$ (Giesekus) and ($\xi$, $\epsilon$)=(0,0.4) for (ePTT).}
	\label{fig:plot_5}
\end{figure}

We notice that the UCM and Giesekus cases exhibited a more elliptical shape of the viscous Lissajous curves. In contrast, the Johnson-Segalman and ePTT cases have shown a more pronounced deviation from the elliptical behavior. As analyzed theoretically by \cite{SouzaMendes2013, Thompson2015} and confirmed experimentally by \cite{SouzaMendes2014,SouzaMendes2018} an elliptical output has one of two reasons: (i) the material is subjected to a small amplitude oscillatory shear (SAOS) process or; (ii) the material is in quasi-linear large amplitude oscillatory shear, QL-LAOS. While most of the readers are familiar with the first process, the second one is associated with a constant structure state, where even out of the linear viscoelastic regime, the material does not have time for structure changes within a cycle when the frequency is high enough. To illustrate the above rationale, we plotted the viscous-Lissajous curves of the Johnson-Segalman and ePTT models for higher Deborah numbers, leading to elliptical trajectories shown in Figure \ref{fig:frequency_JS_PTT}. This result is an important finding. It reveals that nonlinear viscoelastic models have an underlying structure from the perspective of the baseline Maxwell model, something already shown for thixotropic elasto-viscoplastic materials but not yet found in purely viscoelastic ones.

\begin{figure}[H]
    \centering
	\subfloat[Johson-Segalman model]{\includegraphics[width=0.85\linewidth]{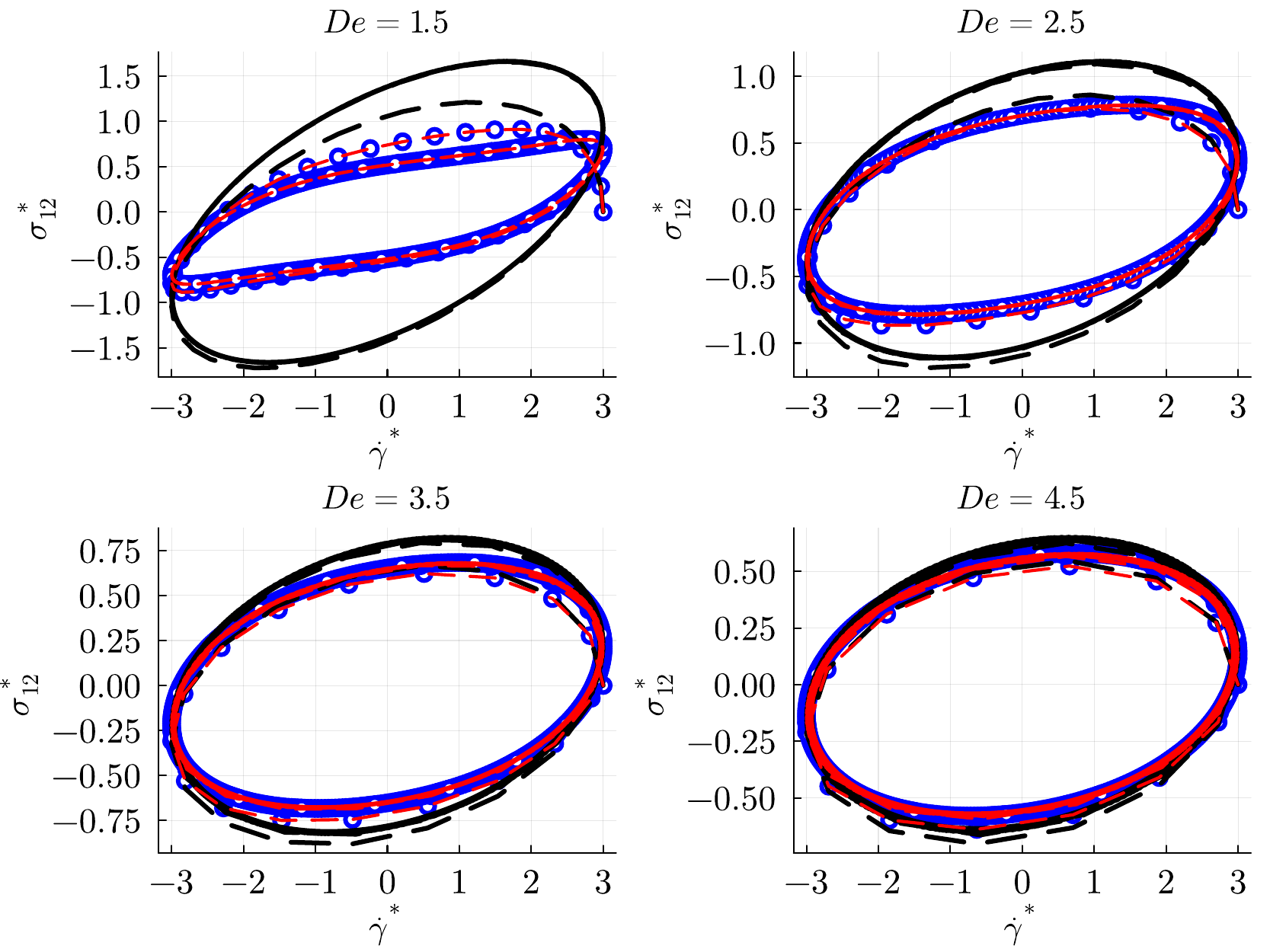}}\label{fig:subfig_JS}\\
     \subfloat[ePTT model]{\includegraphics[width=0.85\linewidth]{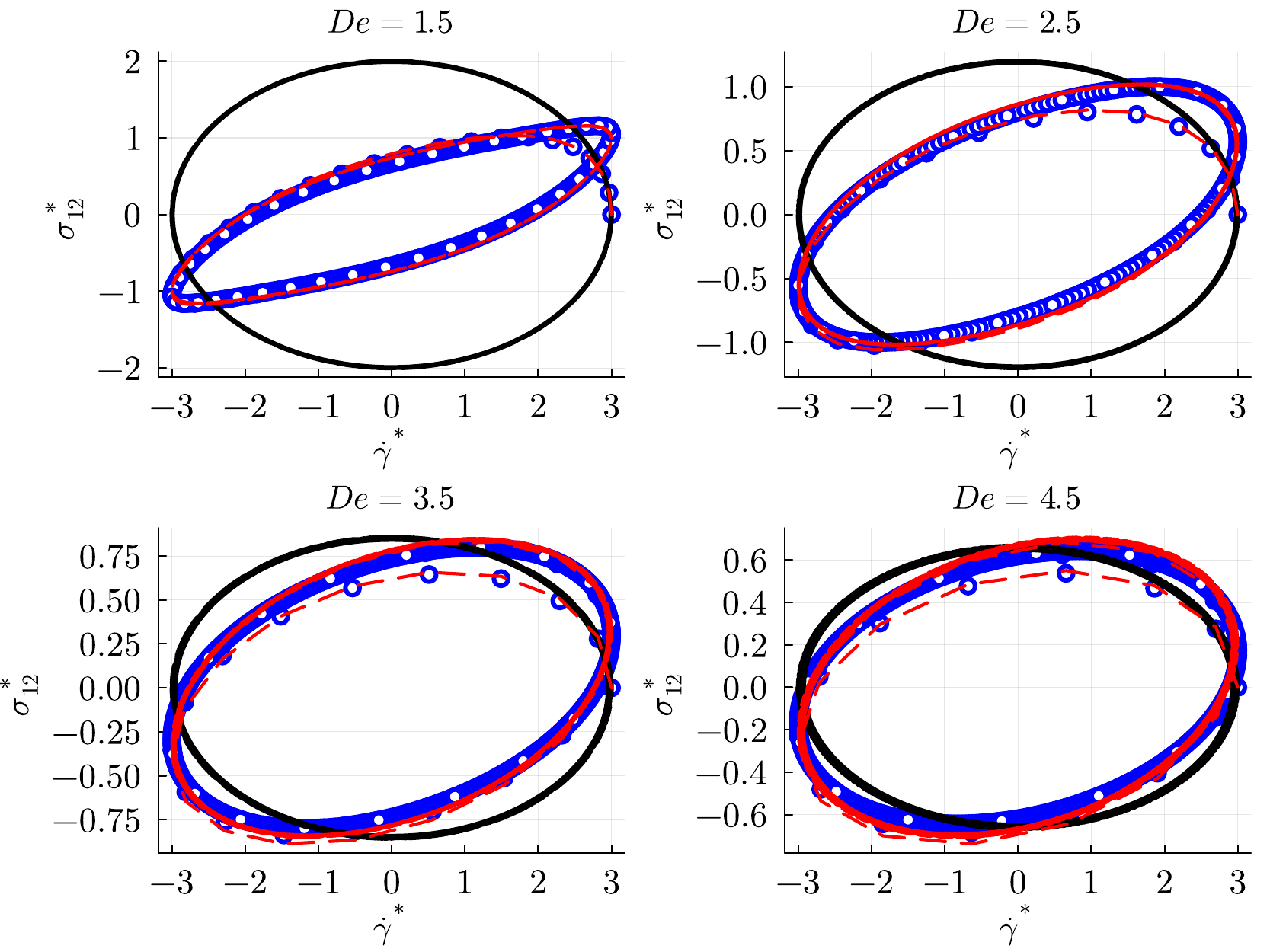}\label{fig:subfig_PTT}}
     \caption{Lissajous-Bowditch curve for different Deborah numbers with input $\dot\gamma^*(t^*)=$ 3\,cos\,(De\,$t^*$).}
    \label{fig:frequency_JS_PTT}
\end{figure}

\begin{figure}
	\centering
	\includegraphics[width=0.9\linewidth]{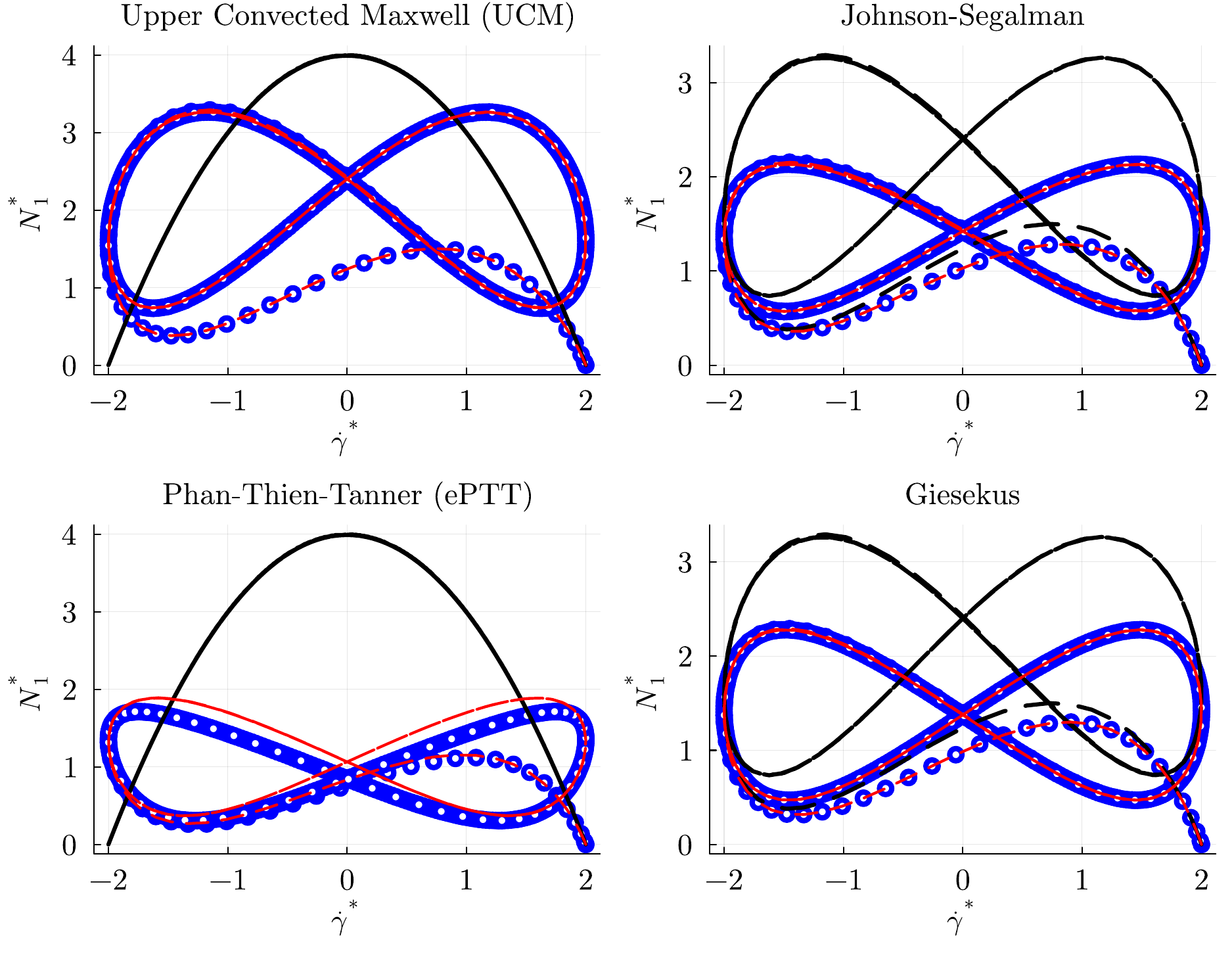}
	\caption{Lissajous-Bowditch curves for the extrapolation of UDE models for the normal stress difference ($N^{*}_1$,${\dot{\gamma}}^*$) with input $\dot\gamma^*(t^*)=$ 2\,cos\,($t^*$). The blue curve illustrates the numerical solution of the differential equation, referred to as the ground truth solution. The black curve depicts the UDE model pre-training, while the red curve illustrates the predicted UDE model. The following parameters were used $\xi=0.4$ (Johnson-Segalman), $\alpha=0.2$ (Giesekus) and ($\xi$, $\epsilon$)=(0,0.4) for (ePTT).}
	\label{fig:plot_6}
\end{figure}

\begin{figure}
	\centering
	\includegraphics[width=0.9\linewidth]{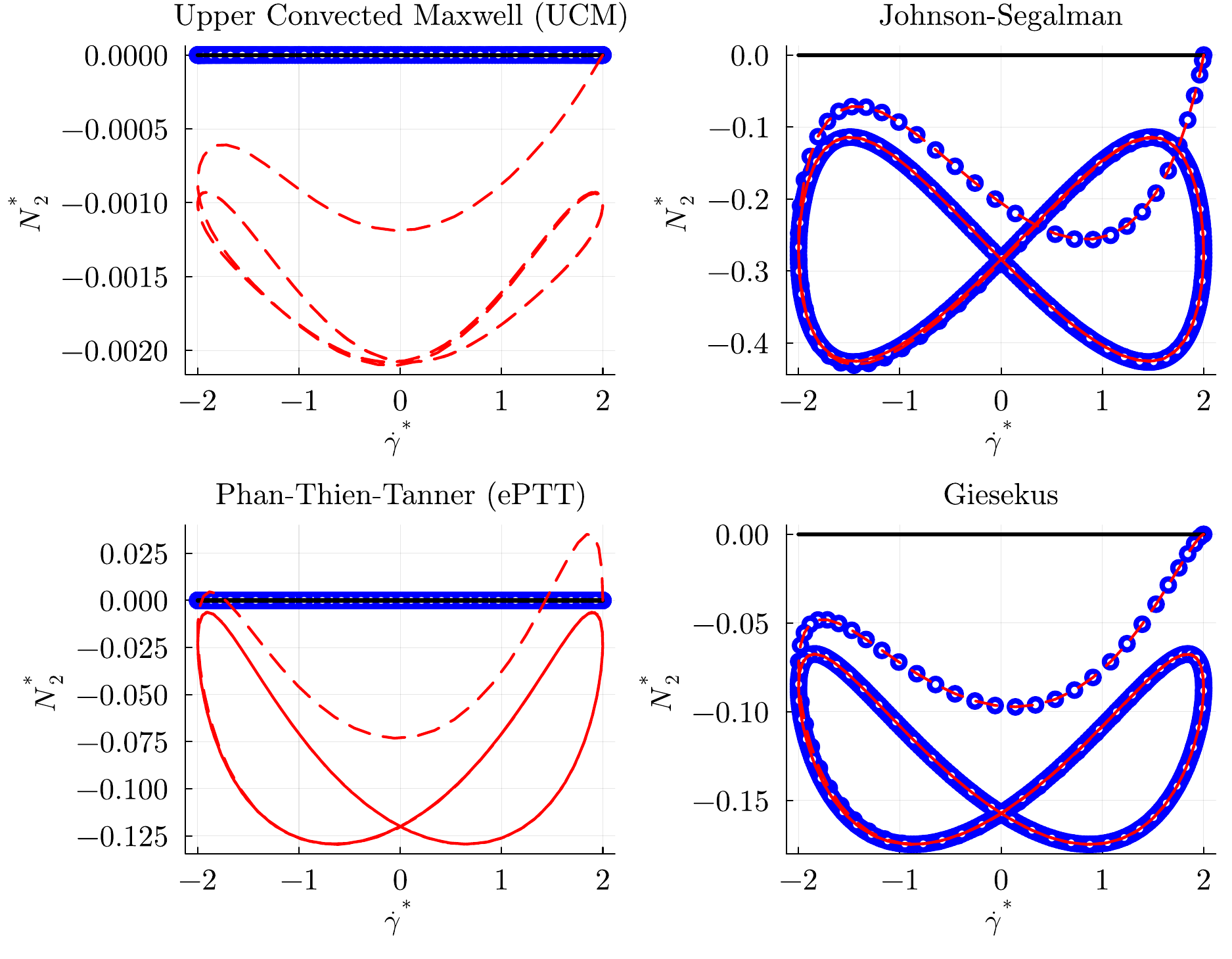}
	\caption{Lissajous-Bowditch curves for the extrapolation of UDE models for normal stress difference ($N^{*}_2$,${\dot{\gamma}}^*$) with input $\dot\gamma^*(t^*)=$ 2\,cos\,($t^*$). The blue curve illustrates the numerical solution of the differential equation, referred to as the "ground truth" solution. The black curve depicts the UDE model pre-training, while the red curve illustrates the UDE model post-training. The following parameters were used $\xi=0.4$ (Johnson-Segalman), $\alpha=0.2$ (Giesekus) and ($\xi$, $\epsilon$)=(0,0.4) for (ePTT).}
	\label{fig:plot_7}
\end{figure}

The curves in the space defined by the first normal stresses N$^{*}_1$ and the shear rate $\dot{\gamma}$ can be observed in Figure \ref{fig:plot_6}. As expected, the black lines of UCM and ePTT exhibited a parabolic shape associated with a purely elastic material having constant first normal stress coefficient in shear, $\psi_1=2\eta\lambda$, which in turn leads to a quadratic behavior of $N_1=\psi_1\dot{\gamma}^2$. In addition, the predicted UCM model coincides with the black lines of the Giesekus and Johnson-Segalman cases. The observed behavior corresponds to a Lemniscate, meaning a non-unitary frequency ratio and a phase discrepancy of $\pi/4$. The model aligns with the ground truth solution, except for the ePTT model, which shows a slight discrepancy. Figure \ref{fig:plot_7} displays the second normal stress difference, N$^{*}_2$. It illustrates how the model accurately represents the behavior seen in the Giesekus and Johnson-Segalman models. Analogous to the prediction of (N$^{*}_2$,$t^*$) in Figure \ref{fig:plot_3}, the model trained with data from the UCM generates negligible prediction values for N$^{*}_2$, once the ratio N$^{*}_2$/N$^{*}_1$ is close to zero.

As for the predictions made by the UDE model, trained with the data generated by the ePTT model, the response of the coefficients is quite complex with $g^{(1)}$ of the order of $10^{-7}$, $g^{(6)}$ of $10^{-5}$, $g^{(2)}\approx 1.11$,  $(g^{(3)},g^{(5)})\approx -0.04$, $g^{(4)}\approx 0.246$ and $(g^{(7)},g^{(8)})\approx -0.02$. 
Decomposing the PTT model into tensor-based functions is challenging since it involves an exponential function. As a result, most of the tensor bases were essential to incorporate the exponential behavior. For this reason, we found non-vanishing values for N$^{*}_2$ in Figure \ref{fig:plot_3} and \ref{fig:plot_7} since the coefficient $g^{(4)}\approx 0.246$ is associated with the tensor ${\mathbf{T}_{ij}^{*}}^{(4)} = \bm{\sigma^*}\boldsymbol\cdot\bm{\sigma^*}$ that corresponds to a Giesekus response (which predicts non-zero values for N$^{*}_2$).} 

\begin{figure}
	\centering
	\includegraphics[width=0.6\linewidth]{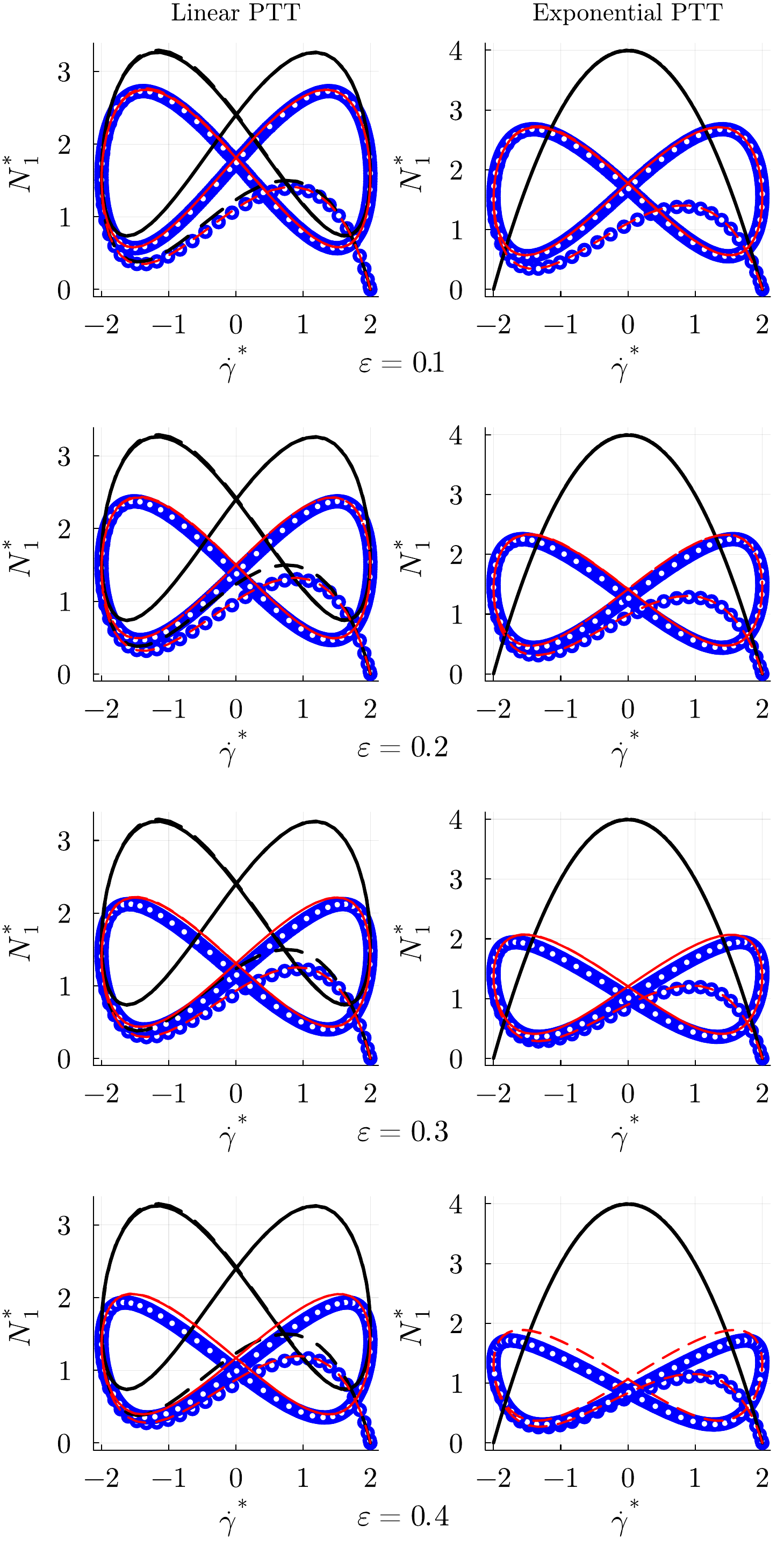}
	\caption{Comparison between UDE prediction for Linear and Exponential PTT models for different extensibility parameters $\epsilon$.}
	\label{fig:plot_PTT}
\end{figure}

It is interesting to note that the recovery response becomes rather complex as the 
extensibility parameter ($\epsilon$) increases. In Figure \ref{fig:plot_PTT}, the predictions of UDE are illustrated for both linear and exponential PTT models across various extensibility parameters ($\epsilon$). In the context of the linear PTT model, the function $h(\bm\sigma)$ presented in Table \ref{table:model} is identified as $h(\bm\sigma)=\dfrac{\sigma}{\lambda}\dfrac{\epsilon}{G}tr(\bm\sigma)$. It is worth mentioning that when $\epsilon$ increases from 0.1 to 0.4, the ability of the PTT models to make accurate predictions decreases, suggesting that a more sophisticated tensorial basis approach is required to represent the behavior accurately.

\begin{figure}
	\centering
	\includegraphics[width=0.95\linewidth]{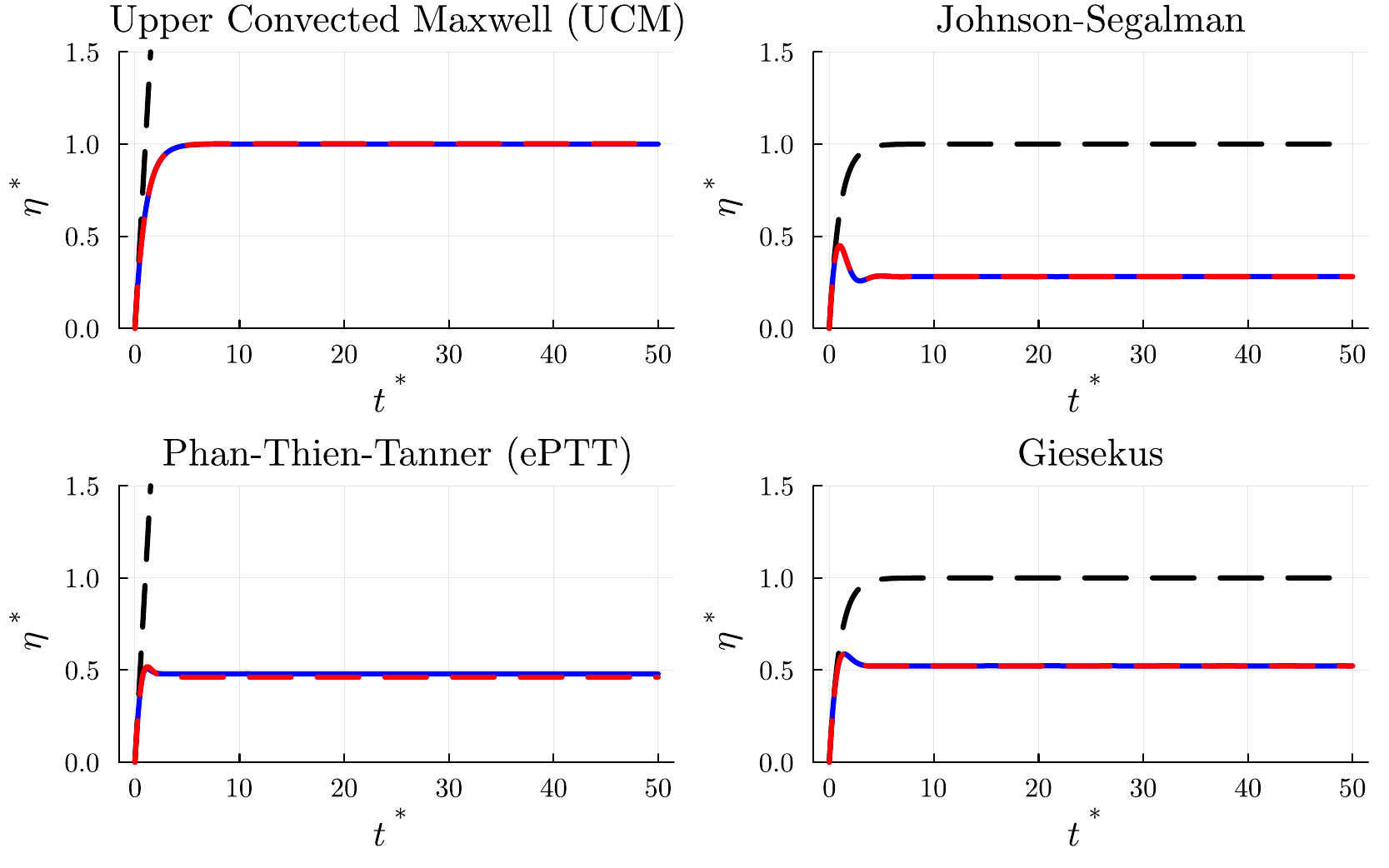}
	\caption{Test for predicting UDE models in a startup experiment with $\dot\gamma^*(t^*)=$ 2.  The blue curve illustrates the numerical solution of the differential equation, referred to as the "ground truth" solution. The black curve depicts the UDE model pre-training, while the red curve illustrates the UDE model post-training. The following parameters were used $\xi=0.4$ (Johnson-Segalman), $\alpha=0.2$ (Giesekus) and ($\xi$, $\epsilon$)=(0,0.4) for (ePTT).}
	\label{fig:plot_4}
\end{figure}

As another test to predict a flow pattern different from the one it was trained on, a startup flow scenario with a shear rate of 2 is exhibited in Fig.~\ref{fig:plot_4}. Once again, the method accurately represents the qualitative trends of the dimensionless viscosity for all models, showing a tinny deviation for the ePTT model. The findings indicate that the methodology's effectiveness extends to various situations, given that the predictions were accurate in a test with different characteristics than the training data.

\newpage

\subsection{Viscoelastic model distillation}

In machine learning, knowledge distillation \cite{Hinton2015,Gou2021,Adsera2024} involves exchanging a complex model (teacher model) with a simpler one (student model) that closely mimics the original one, serving as a form of model compression \cite{Bucilua2006}. Neural network models with elaborate architectures comprising multiple layers and model parameters are more frequently subjected to knowledge distillation. Most distillation techniques are based on capturing the knowledge at some point in the teacher's neural network architecture, i.e., the intermediate or output layer of the neural network. 

Within the concept of distillation, instead of manipulating the network's layers output, we used a distinct strategy to help a basic model extract knowledge from a complex model by training it with a dataset from a complex nonlinear model. The main idea behind this strategy is to adjust the parameters of the basic model to mimic the nonlinear model, making the simpler model a surrogate for the complex one. This enables an unbiased comparison of varying viscoelastic models.

The first step was to analyze the most suitable Upper Convected Maxwell model that could emulate the results generated by the Giesekus model. The neural network was adjusted to generate solely $g^{(2)}$ and $g^{(3)}$ coefficients to recover the UCM model outlined in equation (\ref{eq:surrogate-MG}).

\begin{equation}
    \centering
    \dfrac{d\bm\sigma^*}{dt^*} =(\bm\nabla {\bm{v})^*}^T\boldsymbol\cdot \bm{\sigma^*}+\bm{\sigma^*}\boldsymbol\cdot(\bm\nabla\bm{v})^*-g^{(2)}\bm{\sigma}^*+g^{(3)}{\bm{\dot\gamma}}^*
	\label{eq:surrogate-MG}
\end{equation}

\begin{figure}[H]
    \centering
    \subfloat[]{\includegraphics[width=0.4\linewidth]{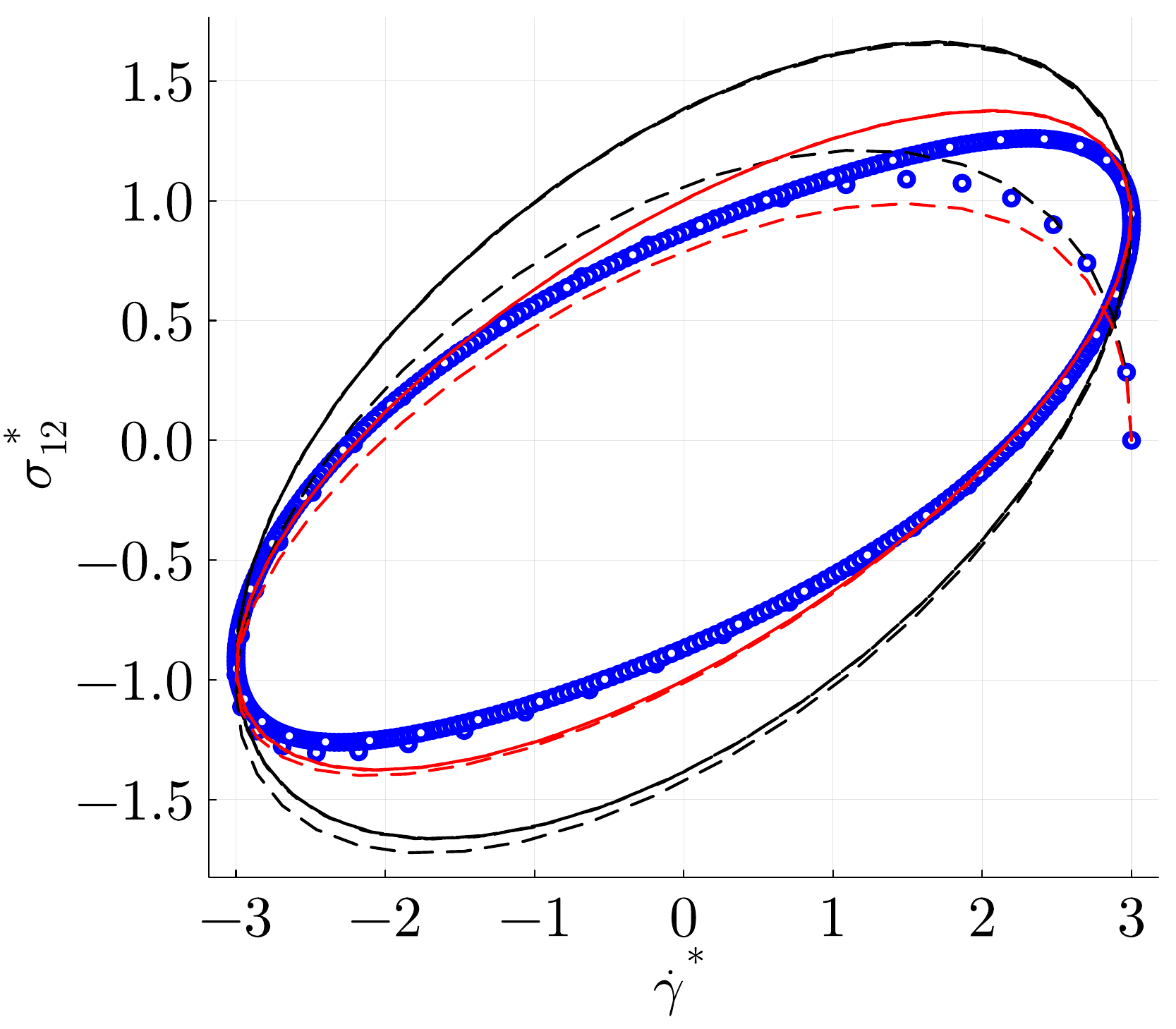}\label{fig:subfig1}} 
    \subfloat[]{\includegraphics[width=0.4\linewidth]{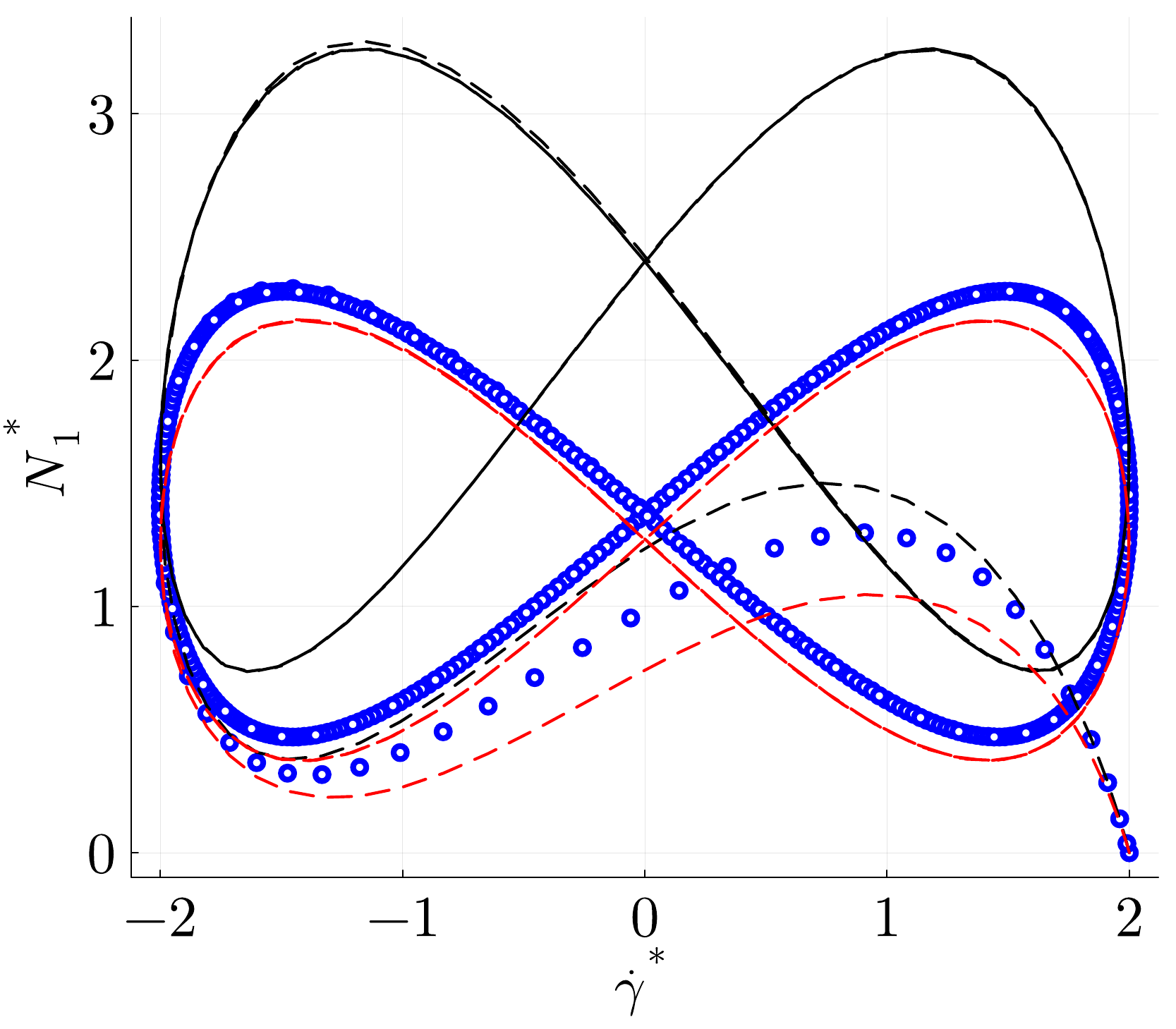}\label{fig:subfig2}}\\  
    \subfloat[]{\includegraphics[width=0.4\linewidth]{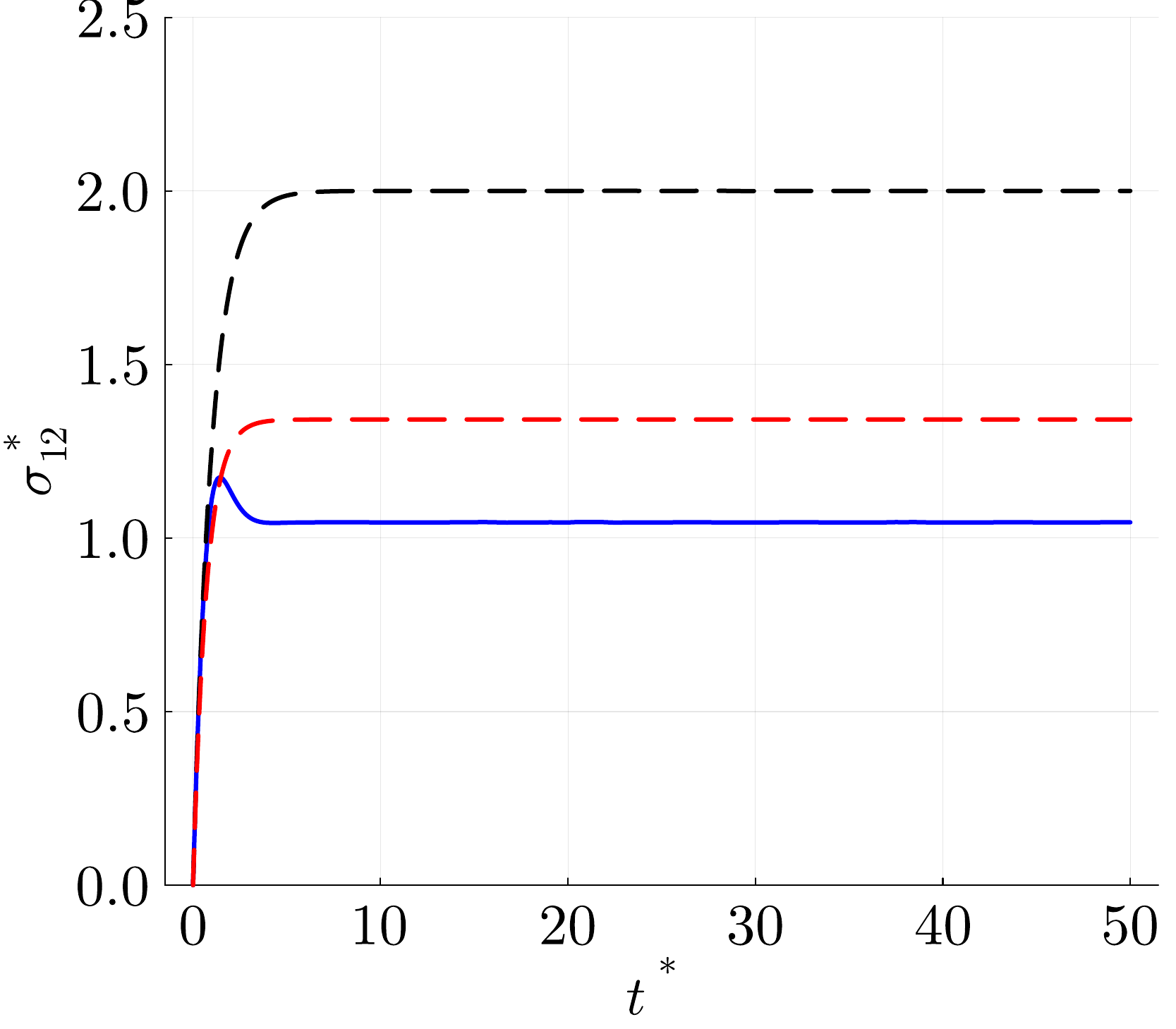}\label{fig:subfig3}}
    \caption{Surrogate Maxwell for Giesekus model. The red curve illustrates the surrogate model post-training. The black curve represents the original UCM model and the blue curve is the "ground truth" solution of the Giesekus model. The recovering coefficients for surrogate Maxwell were $g^{(2)}=1.41$ and $g^{(3)}=0.94$. The training dataset from the Giesekus model was generated with $\alpha = 0.2$. (a) Lissajous-Bowditch with  input $\dot\gamma^*(t^*)=$ 3\,cos\,(1.5\,$t^*$); (b) Lissajous-Bowditch curve for the first normal stress difference N$^*_1$ with input $\dot\gamma^*(t^*)=$ 2\,cos\,($t^*$); (c) Startup experiment with input $\dot\gamma^*(t^*)=$ 2.}
    \label{fig:surrogate_Maxwell_Giesekus}
\end{figure}

Afterward, the coefficients in equation (\ref{eq:surrogate-MG}) were fine-tuned with a dataset produced by the Giesekus model with $\alpha=0.2$. Upon completion of the optimization, it was found that the most favorable coefficients were $g^{(2)}=1.408$ and $g^{(3)}=0.944$. Figure \ref{fig:surrogate_Maxwell_Giesekus} shows the test results concerning the Maxwell surrogate model. The shear stress extrapolation in Figure \ref{fig:subfig1} closely resembles the Giesekus model, with only slight variations in amplitude. On the other hand, the black curves, associated with the original Maxwell model, i.e. simply setting $\alpha=0$ in the Giesekus model, led to a more discrepant output.

Regarding the first normal stress difference, Figure \ref{fig:subfig2}, the distinction between the surrogate Maxwell and Giesekus responses is more prominent in the initial times. Still, the periodic solution has shown the same trends obtained in the shear component case, i.e. the new Maxwell coefficients could capture the essence of $N_1$ behavior. For the start-up experiment, Figure \ref{fig:subfig3}, the distinction is more evident, once the UCM model cannot capture overshooting. Despite this fact, we notice a substantial approximation of the surrogate Maxwell model to the Giesekus solution compared to the original Maxwell model. Given the model's simplicity, the results depicted in Fig.\ref{fig:surrogate_Maxwell_Giesekus} highlight an important point.

These results show that to linearize the nonlinear behavior of the Giesekus model resulted from the inclusion of the quadratic term in the stress, from the perspective of the Maxwell equation, new coefficients associated with $\dot{\gamma}$ and $\sigma$ are necessary. This fact was explored by \cite{Thompson2021,Figueiredo2024} in the context of dimensionless numbers in viscoelastic flows. In that case, they employed the shear rate at the wall of a fully developed part of the domain as a characteristic quantity to choose the parameters of the simple model. Herein, the process of optimization throughout the training phase determined the associated coefficients. To provide a physical interpretation for these new coefficients, $g^{(2)}=1.408$ and $g^{(3)}=0.944$ found in the method, we can revisit the dimensionless analysis and rewrite the viscosity and relaxation time of the surrogated Maxwell model as a function of $\eta$ and $\lambda$ associated with the Giesekus model. This analysis shows that

\begin{figure}[H]
	\centering
	\subfloat[]{\includegraphics[width=0.4\linewidth]{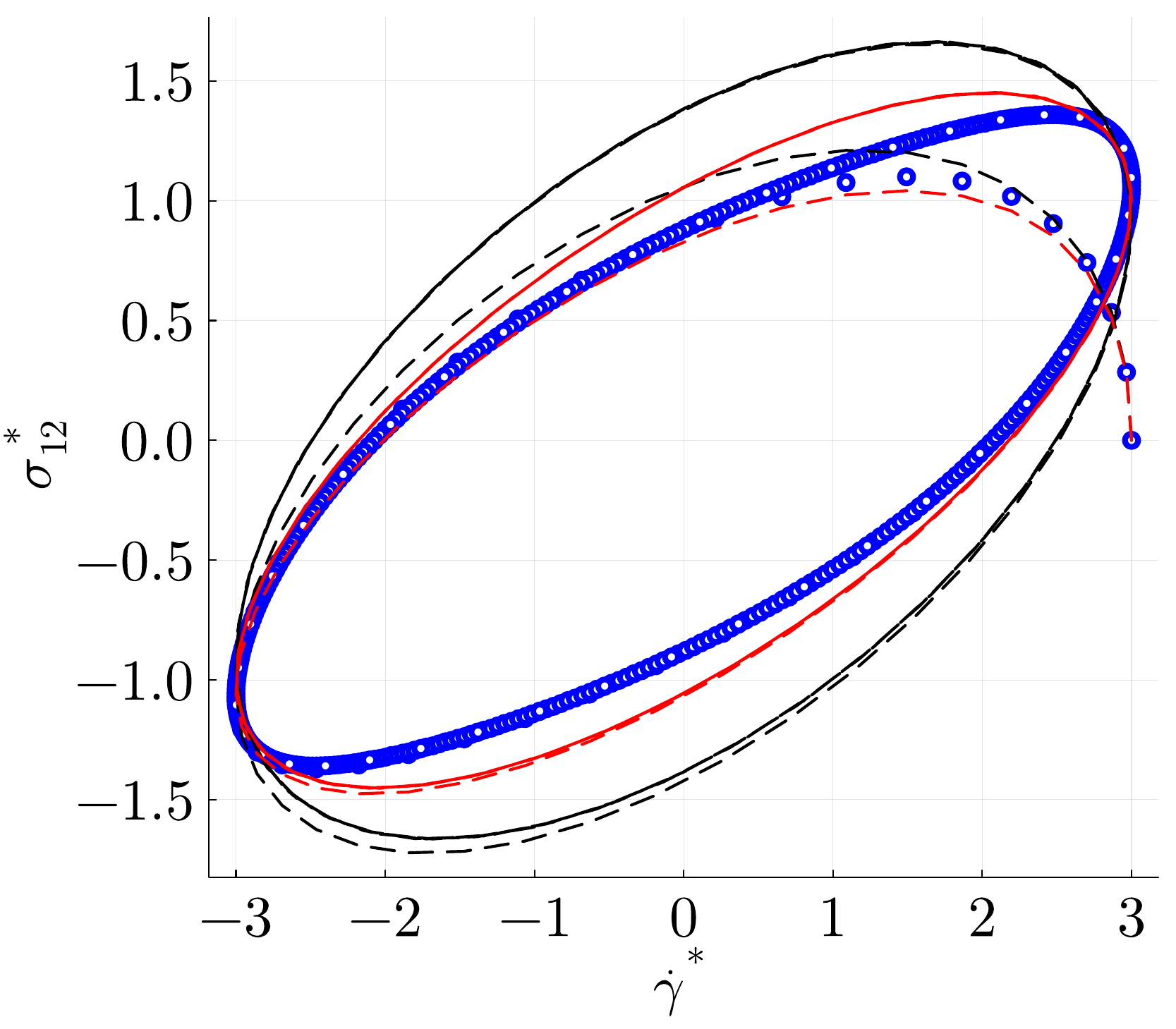}\label{fig:subfig4}}
	\subfloat[]{\includegraphics[width=0.4\linewidth]{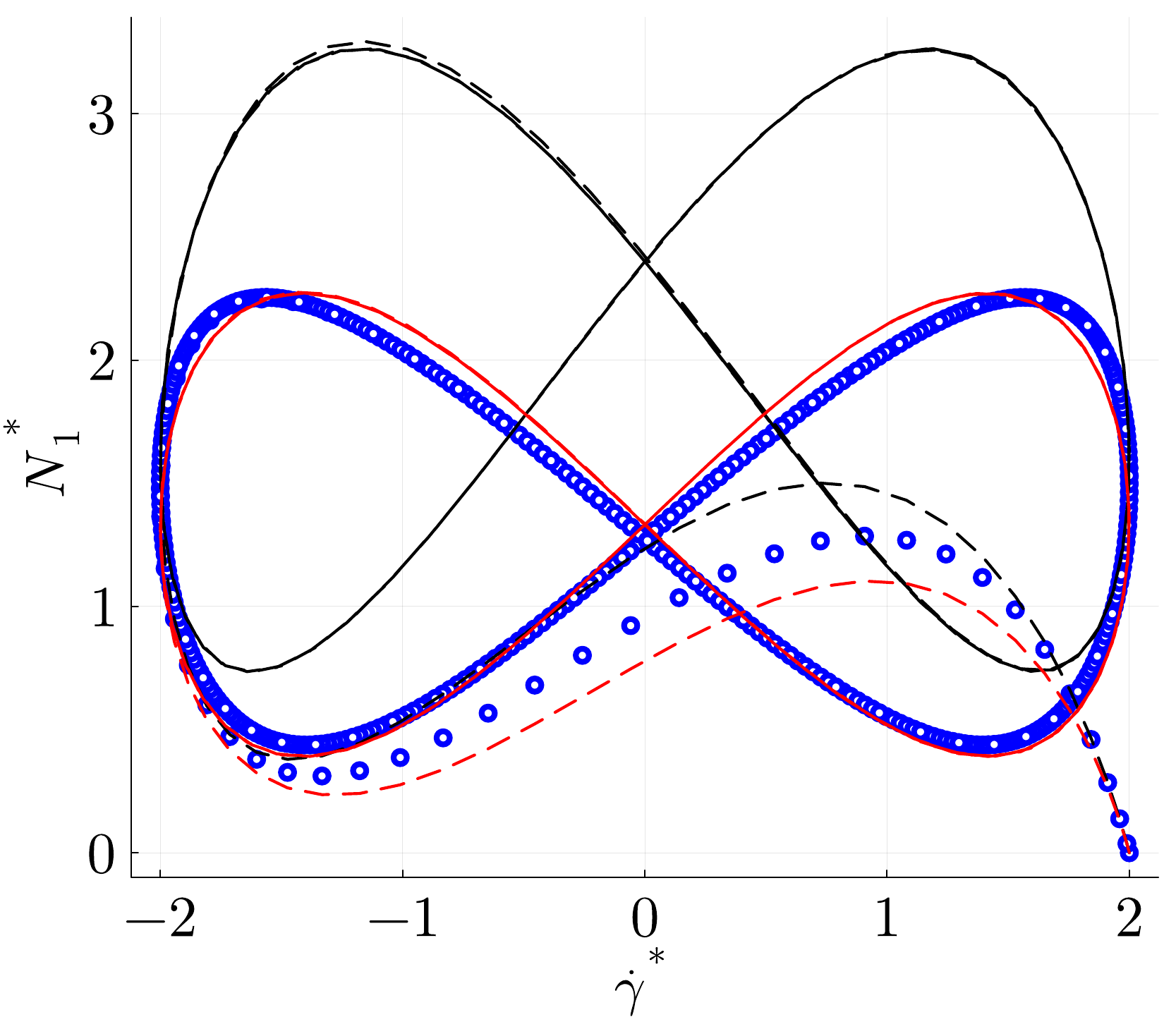}\label{fig:subfig6}}\\
	\subfloat[]{\includegraphics[width=0.4\linewidth]{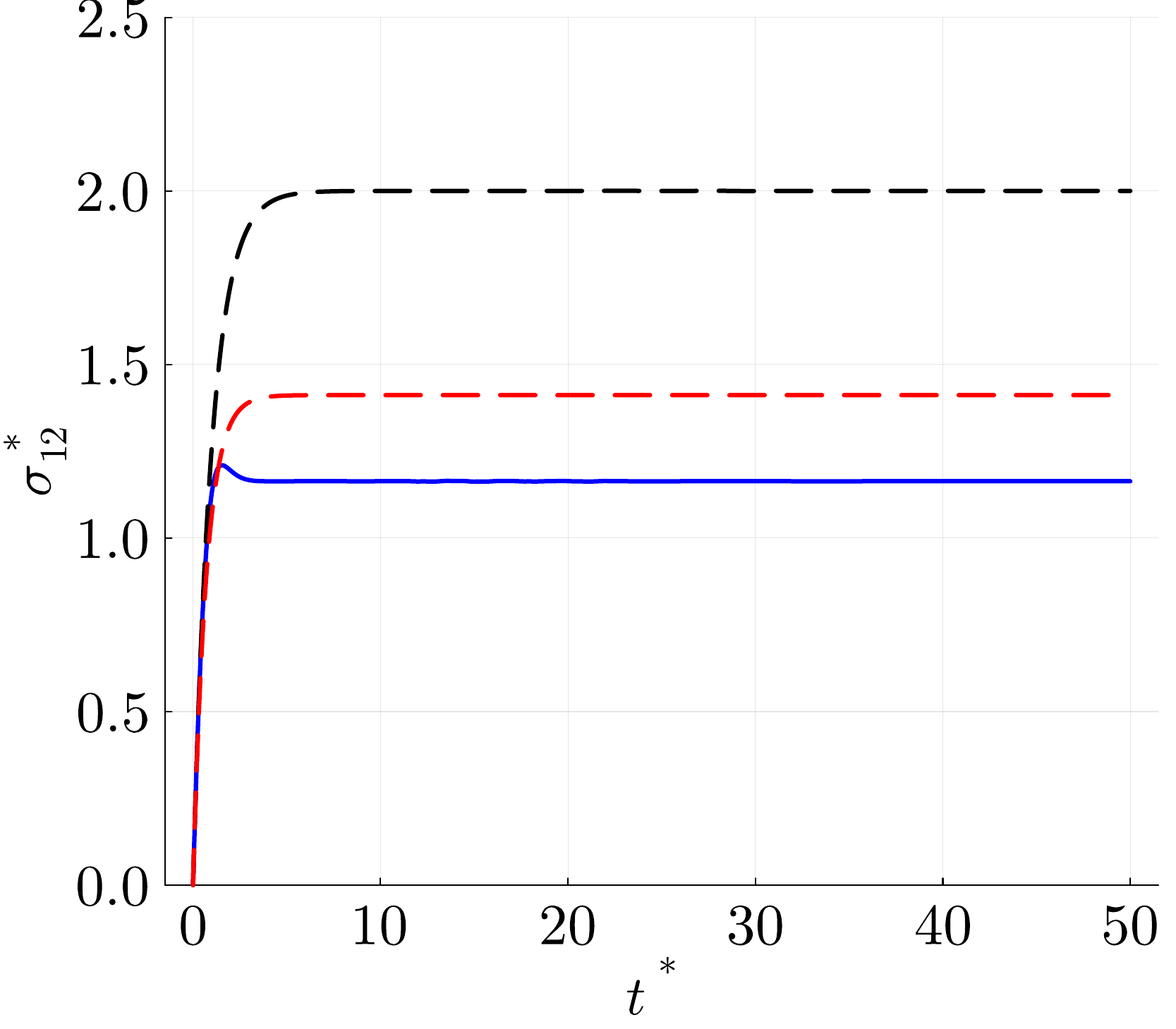}\label{fig:subfig5}}  
    \caption{Surrogate Maxwell for ePTT model. The red curve illustrates the surrogate model post-training. The black curve represents the original UCM model and the blue curve is the "ground truth" solution of the ePTT model. The recovering coefficients for surrogate Maxwell were $g^{(2)}=1.413$ and $g^{(3)}=0.997$. The training dataset from the ePTT model was generated with $(\xi,\epsilon) = (0,0.2)$. (a) Lissajous-Bowditch curve with input $\dot\gamma^*(t^*)=$ 3\,cos\,(1.5\,$t^*$); (b)  Lissajous-Bowditch curve for the first normal stress difference N$^*_1$ with input $\dot\gamma^*(t^*)=$ 2\,cos\,($t^*$); (c) Startup experiment with input $\dot\gamma^*(t^*)=$ 2.}
    \label{fig:surrogate_Maxwell_ePTT}
\end{figure}

\begin{align}
     \eta_{MW}=\dfrac{g^{(3)}}{g^{(2)}}\eta_{G} = 0.670\, \eta_{G}\\
     \label{eq:eta_adjust_G}
     \lambda_{MW}=\dfrac{1}{g^{(2)}}\lambda_{G} = 0.710\, \lambda_{G}\\
     \label{eq:lambda_adjust_G}    
\end{align}

where the subscripts \emph{MW} and \emph{G} refer to Maxwell and Giesekus, respectively. This result is consistent with the behavior of the Giesekus model. The viscosity function in this case is shear-thinning with the zero shear rate (ZSR) value corresponding to $\eta_{G}$. Hence, it is expected that the characteristic viscosity of the problem is below its ZSR value, and by consequence, $\eta_{MW}<\eta_G$. Similarly, the function $\lambda(\dot{\gamma})=\frac{\psi_1(\dot{\gamma})}{2\eta(\dot{\gamma})}$ is also a decreasing function of the shear rate, and hence it is expected that its effective $\lambda$ is lower than the nominal one, $\lambda_{G}$.

The same procedure was adopted using the UCM model as a surrogate for the Exponential Phan-Thien-Tanner (ePTT). The training data set was generated from the ePTT model with $\xi=0$ and $\epsilon=0.2$. The corresponding coefficients were $g^{(2)}=1.413$ and $g^{(3)}=0.997$ and the Maxwell physical parameters present a relation with respect to the ePTT nominal parameters given by

\begin{align}
         \eta_{MW}=\dfrac{g^{(3)}}{g^{(2)}}\eta_{ePTT} = 0.706\, \eta_{ePTT}  \label{eq:eta_adjust_PTT}\\
     \lambda_{MW}=\dfrac{1}{g^{(2)}}\, \lambda_{ePTT} = 0.708\lambda_{ePTT} \label{eq:lambda_adjust_PTT} 
\end{align}

where the subscript \emph{ePTT} stands for the ePTT model. Figure  \ref{fig:surrogate_Maxwell_ePTT} shows that the same tendency of the Maxwell-Giesekus analysis was obtained with the Maxwell-ePTT analysis for the three quantities, namely oscillatory shear stress, oscillatory first normal stress difference, and start-up shear stress. The surrogated Maxwell model could closely mimic the ePTT model, with minor deviations in amplitude in the oscillatory cases and a more pronounced difference in the start-up case. The discrepancy between original Maxwell and surrogated Maxwell is remarkable in all cases.  

It is worth noticing that the usual procedure in the literature adopted to compare nonlinear and linear models is to evaluate the discrepancy between the blue and black curves, as a representation of the nonlinear effect associated with a non-zero value of $\alpha$ or $\epsilon$. The procedure outlined here, obtained by generating the red curves, offers another view where a part of the nonlinear effect can be absorbed in the linear description if the parameters of the linear model are judiciously chosen. Herein, \emph{judiciously chosen} parameters are determined during the optimization process of the Universal Differential Equation procedure that optimizes the linear representation of the nonlinear model.

\section{Discussion and Outlook}

We have shown the use of differentiable physics from universal differential equations for the physics-informed data-driven modeling of non-Newtonian constitutive equations. A differential equation describes part of the behavior of a viscoelastic fluid, working as prior physical knowledge for the constitutive equation. A tensor basis feed-forward neural network (TBNN) embedded into the differential equation captures missing terms in the constitutive equation, resulting in a hybrid formulation called Universal Differential Equation (UDE).

Four viscoelastic models, Upper Convected Maxwell (UCM), Giesekus, Johnson-Segalman, and Exponential Phan-Thien-Tanner (ePTT), were employed to examine the methodology. Through eight distinct oscillatory experiments, a synthetic dataset for training was created, consisting of 32 time series ($\sigma_{12}^*$, $t^*$) to represent shear stress only. For each model, terms in the constitutive equation were considered missing so that TBNN could recover them. 

Regardless of the synthetic data generation model utilized, the universal differential equation is capable of capturing patterns beyond the training data for normal stress differences, whether in terms of time, amplitude, or frequency. In the case of the ePTT model, the recovery behavior was less accurate due to the embedded exponential function and an evolving parameter as a function of $tr(\bm\sigma^*)$. These results showed that future modifications to the network are necessary to deal with parameters as functions or even to use other types of networks and tensorial basis. Nevertheless, the approach holds the potential for handling more sophisticated models.

The analysis associated with the LAOS procedure indicated an important finding, namely the achievement of a quasi-linear regime, referred to QL-LAOS, where the output is sinusoidal and the material functions are reduced to similar versions of the SAOS approach.

In the framework of model distillation, we conducted an experiment in which a simple linear model extracts knowledge from a nonlinear one by training it with a dataset from the more complex model. In this regard, we evaluated how the optimal Upper Convected Maxwell model 
could replicate the outcomes produced by the Giesekus and Exponential Phan-Thien-Tanner models. We found that the discrepancy between nonlinear and UCM models can be substantially reduced when compared to the usual analysis, where the linear model is obtained by vanishing the nonlinear term keeping the other parameters fixed. The optimization procedure adopted here could be interpreted from the physical viewpoint as a change in the viscous and relaxation time parameters of the corresponding linear model that could absorb part of the nonlinear effects. 
These findings indicate a potential use of the method described here by replicating a sophisticated model using a simpler one. One potential use of this method could involve employing a basic parameter model to represent a response with multiple modes. Another potential use of this analysis is to translate one model into another, by taking two models of similar complexity, like Giesekus, linear-PTT, FENE-P (finitely extensible nonlinear elastic with Peterlin closure), and see how the parameters of one model can be optimized by training with data from another model.

\end{document}